\documentclass[preprint]{imsart}
\usepackage[OT1]{fontenc}
\usepackage{amsthm,amsmath}
\usepackage[colorlinks,citecolor=blue,urlcolor=blue]{hyperref}
\usepackage{amsmath,amssymb,amsthm}
\usepackage{mathrsfs,euscript}
\usepackage[english]{babel}
\usepackage[numbers]{natbib}
\usepackage[dvips]{graphicx}
\usepackage{color}
\usepackage{indentfirst}
\usepackage{multirow}
\usepackage{float,bm}
\usepackage{a4wide}

\startlocaldefs
\numberwithin{equation}{section}
\theoremstyle{plain}
\def\@journal{Submitted}
\endlocaldefs

\begin{document}

\newcommand\E{\mathbb{E}}
\newcommand{\cov}{\mathop{\text{Cov}}}
\newcommand\var{\mathop{\text{Var}}}
\newcommand\tr{\mathop{\text{tr}}}
\newcommand\topii{{\frac1{2\pi i}}}
\newcommand\mbar{\underline{m}}
\newcommand\EN{\EuScript{N}}
\newcommand\cvd{\stackrel{\EuScript{D}}{\longrightarrow}}
\newcommand{\diag}{{\rm diag}}
\newcommand{\sjln}{\sum_{j=1}^n}
\newcommand{\bqn}{\begin{eqnarray*}}\newcommand{\eqn}{\end{eqnarray*}}
\newcommand{\bqa}{\begin{eqnarray}}\newcommand{\eqa}{\end{eqnarray}}
\newcommand{\um}{{\underline{m}}}
\newcommand{\cC}{{\mathcal C}}
\newcommand{\bA}{\mathbf{A}}\newcommand{\bB}{\mathbf{B}}
\newcommand{\bF}{\mathbf{F}}\newcommand{\bG}{\mathbf{G}}
\newcommand{\bS}{\mathbf{S}}\newcommand{\bT}{\mathbf{T}}
\newcommand{\bX}{\mathbf{X}}\newcommand{\bY}{\mathbf{Y}}
\newcommand{\bZ}{\mathbf{Z}}\newcommand{\bI}{\mathbf{I}}
\newcommand{\bbx}{\mathbf{x}}\newcommand{\bby}{\mathbf{y}}
\newcommand{\bmu}{\boldsymbol{\mu}}\newcommand{\bnu}{\boldsymbol{\nu}}
\newcommand{\bSi}{\mathbf{\Sigma}}
\newcommand{\bGa}{\mathbf{\Gamma}}
\newcommand{\CC}{{\mathbb{C}}}
\newcommand{\bgma}{\boldsymbol{\gamma}}
\newcommand{\gD}{\Delta}\newcommand{\gd}{\delta}
\newcommand{\bbe}{{\bf e}}\newcommand{\bbA}{{\bf A}}
\newcommand{\bbB}{{\bf B}}\newcommand{\bbS}{{\bf S}}
\newcommand{\bbT}{{\bf T}}\newcommand{\bbI}{{\bf I}}
\newcommand{\bbr}{\mathbf{r}}
\newcommand{\rtr}{{\rm tr}}
\newcommand{\rE}{{\rm E}}
\newcommand{\ointctrclockwise}{\oint}
\newcommand{\cU}{{\cal U}}
\newcommand{\rdd}{\textcolor{red}}
\newcommand{\eprf}{\hspace*{\fill}$\blacksquare$}
\newcommand{\bbU}{\mathbf{U}}
\newcommand{\bgL}{\mathbf{L}}

\newtheorem{thm}{Theorem}[section]
\newtheorem{prop}{Proposition}[section]
\newtheorem{lem}{Lemma}[section]
\newtheorem{rem}{Remark}[section]
\newtheorem{cor}{Corollary}[section]
\newtheorem{example}{Example}[section]

\begin{frontmatter}

\title{Substitution principle for CLT of  linear spectral statistics
  of high-dimensional  sample covariance matrices with   applications to
  hypothesis testing}
  \runtitle{Substitution principle for unbiased sample covariance matrices}
  \thankstext{T1}{S. Zheng  was  partially supported by NSFC grants NSFC-11171058 and NECT-11-0616.}
  \thankstext{T2}{Z. D. Bai  was  partially supported by NSFC grant NSFC-11171057, PCSIRT and Fundamental
Research Funds for the Central Universities..}
  \thankstext{T3}{J. Yao was partially supported by a HKU Start-up fund.}

  \begin{aug}
    \author{\fnms{Shurong} \snm{Zheng}\thanksref{T1}\ead[label=e1]{zhengsr@nenu.edu.cn}}
    \and
    \author{\fnms{Zhidong} \snm{Bai}\thanksref{T2}\ead[label=e2]{baizd@nenu.edu.cn}}
    \and
    \author{\fnms{Jianfeng} \snm{Yao}\thanksref{T3}\ead[label=e3]{jeffyao@hku.hk}}
    \runauthor{S. Zheng,  Z. D. Bai and J.Yao}
    \affiliation{Northeast Normal University and
      The University of Hong Kong}
   \address{Shurong  Zheng and Zhidong Bai   \\
      School of Mathematics $\&$ Statistics and KLAS\\
      Northeast Normal University \\
      Changchun, China\\
      \printead{e1,e2}
    }

    \address{Jianfeng Yao \\
      Department of Statistics and Actuarial Science\\
      The University of Hong Kong\\
      Pokfulam, \quad
      Hong Kong \\
      \printead{e3}
    }
  \end{aug}

  \begin{abstract}
    Sample covariance matrices
     are widely
    used in multivariate statistical analysis.
    The central limit theorems (CLT's) for linear spectral statistics of
     high-dimensional non-centered sample covariance matrices
    have received considerable attention in random matrix theory and
    have been applied to many high-dimensional statistical
    problems. However, known population mean vectors are assumed for non-centered sample covariance matrices, some of which even assume Gaussian-like moment conditions.
    In fact, there are still another two most frequently used sample covariance matrices:
    the MLE (by subtracting the sample mean vector from each sample vector) and the unbiased sample covariance matrix (by changing the denominator $n$ as $N=n-1$ in the MLE)
    without depending on unknown population mean vectors.
    In this paper, we not only establish new CLT's for non-centered sample covariance matrices
    without Gaussian-like moment conditions but also characterize the non-negligible differences among the CLT's for the three classes of 
   high-dimensional sample covariance matrices
    by establishing a {\em substitution principle}: substitute
    the {\em adjusted} sample size $N=n-1$ for the actual sample size $n$ in the major centering term of the new CLT's so as to obtain the CLT of the unbiased sample covariance matrices.
    Moreover, it is found that
    the difference between the CLT's for the MLE and unbiased sample covariance matrix
    is non-negligible
    in the major centering term although the two sample covariance matrices only have differences $n$  and $n-1$ on the dominator.
    The new results are applied to two testing problems for
    high-dimensional data.
\end{abstract}




\end{frontmatter}

\section{Introduction}
Consider a sample ${\bbx}_{1},\cdots,{\bbx}_{ n}$ of size $n$ from
a $p$-dimensional population $\bbx$ with unknown mean $\bmu$ and covariance
matrix $\bSi$. The {\em unbiased sample covariance matrix} is
\begin{equation}
  \bS_n=\frac{1}{N}\sum\limits_{i=1}^n(\bbx_i-\overline{\bbx})(\bbx_i-\overline{\bbx})^*~,
  \label{Sx}
\end{equation}
where $\overline{\bbx}=\frac1n\sum_j \bbx_j$ is the sample mean,
$N:=n-1$ the {\em adjusted sample size} and $*$ denotes transpose and conjugate.
Sample covariance matrices are widely applied in
multivariate statistical analysis.
For example, in structure testing problems of population covariance matrices $\bSi$, many well-known test statistics are functionals of
the eigenvalues
$\{\lambda_j, 1\le j\le p\}$ of $\bS_n$ that have the form
\begin{equation}
  \label{eq:lss}
  T = \frac1p \sum_{j=1}^p g(\lambda_j) =: \mu_{\bS_n}(g)~,
\end{equation}
for some given function $g$. Such
statistics are referred hereafter
as {\em linear spectral statistics} (LSS)  of the unbiased sample covariance matrix
$\bS_n$.
For example, the log-likelihood ratio statistic  for
testing  the identity hypothesis
 for a Gaussian population
is proportional to $\mu_{\bS_n}(g)$ with $g(\lambda)= \lambda-1-\log\lambda$
(see Section~\ref{testing} for more details).
John's test for the sphericity hypothesis
``$\bSi=\sigma^2 {\bf I}_p$'' ($\sigma^2$ unspecified)
uses the square of the coefficient of variation of the sample
eigenvalues
\[
  \label{eq:John}
  U_n =  \frac{p^{-1} \sum_{j=1}^p (\lambda_j-\overline \lambda)^2}{\overline \lambda^2}~,
\]
where $\overline \lambda=p^{-1}\sum_j \lambda_j = \mu_{\bS_n}(\lambda)$.
Clearly, $U_n$ is a function of two linear spectral statistics
$\mu_{\bS_n}(\lambda^2)$ and   $\mu_{\bS_n}(\lambda)$ (see
e.g. \cite{WY13} for more details on this test).
Therefore, LSS
$\mu_{\bS_n}(g)$ of the sample covariance matrix $\bS_n$
are of importance in multivariate analysis.

When the dimension $p$ is much less than the sample size $n$, or
equivalently, the {\em dimension-to-sample ratio} $p/n$ is close to
zero, classical large sample theory assesses that once
$\E\|\bbx\|^4<\infty$,
the sample covariance matrix $\bS_n$ is a consistent and asymptotic normal
estimator of $\bSi$.
Consequently, the same also holds for the sample eigenvalues $\{\lambda_j,j=1,\ldots,p\}$
as an estimator of the population eigenvalues of $\bSi$. Therefore,
\begin{equation}
  \label{eq:classic}
  \sqrt n  \left\{ \mu_{\bS_n} (g) - \mu_{\bSi}(g)\right\} \to \EN(0,s^2)~,
\end{equation}
where the asymptotic variance $s^2$ is a function of
$\bSi$ and $g$.  Here and in all the paper,
$\mu_A:=p^{-1}\sum_i \delta_{\{\alpha_i\}}$ denotes the {\it empirical spectral distribution} (ESD)
generated by the  eigenvalues
$\{\alpha_i, 1\le i\le p\}$ of a matrix $A$,  so that for a given function $g$,
$\mu_A(g) = \frac1p \sum_{i} g(\alpha_i)$.

High-dimensional statistics
have emerged in recent years as an important
and active research area.
Applications have been found in various fields such as
genomic data analysis and wireless communications.
Typically in these problems, the ratio $p/n$ is no more close to zero
and the above large sample theory \eqref{eq:classic} fails to  provide
meaningful inference procedures.
Many efforts have been put in finding new procedures
to deal with high-dimensional data.
As an example, the inconsistency of $\bS_n$ as an estimator of $\bSi$
has lead to an abundant literature on covariance matrix estimation
(see e.g. \citet{BicLev08a,BicLev08b}, \citet{CaiLiu11} and the references
therein).

This paper is concerned with asymptotics of LSS  $\mu_{\bS_n} (g)$.   An
interesting  question is what is  the CLT replacing
\eqref{eq:classic}  in the high-dimensional context?
Notice that it remains challenging to transform
the above-mentioned results on covariance matrix estimation
to limit theorem on LSS of interest.
It turns out that when both the dimension $p$ and the sample size
$n$ grow to infinity, limit theory for sample eigenvalues
depend on how the ratio $p/n$ behaves asymptotically.
In this paper, we adopt the so-called {\em Mar\v{c}enko-Pastur scheme}
where it is assumed that $=p/n\to y \in(0,\infty)$ as $n\to\infty$.
It has been  demonstrated that such limiting scheme has a wide
application scope for real-life high-dimensional data analysis
\citep{Johnstone07}.

The seminal paper
\citet{BS04} establishes such a CLT for
the population with population mean $\bmu=0$ (or equivalently, $\bmu$ is known and
data can then be dealt with by substracting $\bmu$) and Gaussian-like moment conditions (the population 2nd-order and 4th-order moments are the same as those of real or complex Gaussian population),  and the non-centered sample covariance matrix is
defined as
\begin{equation}
  \label{Sx0}
  \bS_n^0=\frac{1}{n}\sum\limits_{i=1}^n \bbx_i^0 {\bbx^0_i}^*~.
\end{equation}
(The superscript $0$ is here to remind the fact the population $\bbx^0$ has zero population mean). Let $y_n:=p/n$ and  $H_p=\mu_{\bSi}$ be the population eigenvalue
distribution  of $\bSi$. As $p,~n\to\infty$, it is
assumed that the  ratio $y_n \to y\in(0,\infty)$
and  $H_p\to H$ (weakly)  for some probability
distribution $H$. Then the ESD
$\mu_{\bS_n^0}$ converges
to a nonrandom distribution $F^{y,H}$, called {\em limiting
spectral distribution} (LSD),  which depends on $y$ and the
population limiting distribution $H$. This LSD is referred as the
{\sl generalized Mar\v{c}enko-Pastur distribution} with index
$(y,H)$ (for background on Mar\v{c}enko-Pastur distributions,
the reader is referred to
\citet[Chapter 3]{BSbook}).
Therefore, in the simplest form, the CLT in \cite{BS04}
states that
\begin{equation}\label{clt0}
  p \left\{ \mu_{\bS_n^0}(g) - F^{y_n,H_p}(g)
  \right\}
  \cvd \EN(m(g),v(g))~,
\end{equation}
a Gaussian distribution whose  parameters $m(g)$ and $v(g)$ depend
only on the LSD $F^{y,H}$ and $g$. The crucial issue here is that
the major centering term  $F^{y_n,H_p}(g):=\int g(x) dF^{y_n,H_p}
(x)$ uses a  finite-horizon proxy of the LSD $F^{y,H}$ obtained by
substituting, in the LSD,   $y_n=p/n$ for $y$ and $H_p$ for $H$,
respectively. These substitutions are necessary because  the
convergence speed is $n$ (or $p\propto  n$)
and  any
mis-estimation of order $n^{-1}$ in   $F^{y_n,H_p}(g)$ will affect
the asymptotic mean $m(g)$.

This scenario of populations with a  known mean $\bmu$,
is however a bit too ideal and
real-life data analyses rely on the unbiased sample covariance matrix
$\bS_n$~\eqref{Sx} after subtraction of the sample mean. It has
been believed for a while in the literature in high-dimensional
statistics that both sample covariance matrices $\bS_n$ and $\bS_n^0$ share a same
CLT for their LSS, i.e. the CLT
\eqref{clt0} might apply equally to the
matrix $\bS_n$. Unfortunately, this is indeed untrue.
The problem
can be best
seen by observing the Gaussian case. Actually, for a Gaussian
population,
$N\bS_n:=\sum\limits_{i=1}^n(\bbx_i-\overline{\bbx})(\bbx_i-\overline{\bbx})^*$
 has a Wishart distribution ${\mathbf W}_{N}(\bSi)$ with $N=n-1$
degrees of freedom. Since from a Gaussian population with known population mean, the
matrix $ N \bS_{N}^0=\sum\limits_{i=1}^{N} \bbx_i^0 {\bbx^0}^* $
has the same Wishart distribution. Then we conclude that the
fluctuations of the eigenvalues $\{\lambda_j\}$ of $\bS_n$  are the
same as the matrix $\bS_{N}^0$ so that by \eqref{clt0}, it holds
\begin{equation}\label{clt1}
  p\left\{ \mu_{\bS_n}(g)   -  F^{{y_{N}},H_p}(g) \right\}
  \cvd \EN(m(g),v(g))~.
\end{equation}
In  words, in the Gaussian case, the CLT for populations with unknown means
is the same as the CLT for populations with known means
provided that in the major centering term $F^{{y_{n}},H_p}(g)$,
one substitutes the
adjusted sample size $N=n-1$ for the sample size $n$.
This result will be referred
hereafter as the {\em substitution principle}. Notice that
typically the difference between $F^{{y_{N}},H_p}(g)$ and
$F^{{y_{n}},H_p}(g)$ is  of order $O(n^{-1})$ and as explained
above, such a difference is non-negligible because of  the
multiplication by  $p$ in the CLT.
  As an example, when
  $\bm{\Sigma}={\bf I}_p$
  $ H_p=\delta_1$ and  for $g(x)=x^2$,
  it is well-known that $F^{y_n,\delta_1}(x^2)=1+y_n$.
  Therefore the difference
  $p\{F^{y_n,H_p}(g)-
  F^{y_N,H_p}(g)\}= p(y_n-y_{N})$ tends to
  $- y^2$, a non-negligible negative constant.%

This substitution principle  is indeed a remarkable result and
provides an elegant solution to the question of CLT for LSS of the
unbiased covariance matrix $\bS_n$ from a Gaussian population. It
then raises the question whether the {principle} is universal,
i.e. valid for general populations other than Gaussian. One  of
the main results from the paper establishes this universality for
arbitrary populations provided the existence of a fourth-order
moment. Meanwhile,  most of the existing
methods in hypothesis testing  or  regression analysis
with high-dimensional data
assume either
Gaussian-like moment conditions or populations with known means, see e.g.
\cite{BJYZ09,BJYZ13,r7,LW02,Sriv05,r26}, so that LSS of the sample
covariance matrices are approximated using either the CLT
\eqref{clt1} or the CLT \eqref{clt0}. The
universality of the substitution principle established
in this paper for these CLT's will then help the existing methods to
cover more general high-dimensional data.
Consider
 the MLE (maximum likelihood estimate) of $\bSi$
  \[ \label{hatSigma}
  \hat{\bSi}_n=\frac{1}{n}\sum\limits_{i=1}^n(\bbx_i-\overline{\bbx})(\bbx_i-\overline{\bbx})^*.
  \]
By the decomposition
$
  \mu_{\hat{\bSi}_n}  -        F^{{{y_{N}}},H_p}
  =\left(  \mu_{\bS_n}  -  F^{{{y_{N}}},H_p} \right)
  +\left( \mu_{\hat{\bSi}_n}  - \mu_{\bS_n} \right),
$
  the CLT \eqref{clt1} established in this paper
  and the fact $\hat{\bSi}_n=(1-1/n)\bS_n$, it readily follows that
 \begin{eqnarray*}
   m_1(g)&=&
   p\left( \mu_{\hat{\bSi}_n}  - \mu_{\bS_n} \right)(g)
  =\sum_{i=1}^p\left\{ g((1-/n)\lambda_i)- g(\lambda_i)\right\}
 \rightarrow
   -y F^{y,H}(xg'(x))~.
 \end{eqnarray*}
That is,
\begin{equation}\label{clt2}
    p \left\{ \mu_{\hat{\bSi}_n}(g) - F^{{{y_{N}}},H_p}(g)\right\}
    + m_1(g) \cvd \EN(m(g),v(g))
  \end{equation}
which shows that the difference of CLT's between
 the MLE and biased sample covariance is non-negligible,
 and that
 the CLT for LSS $\mu_{\hat{\bSi}_n}(g)$ for the MLE $\hat{\bSi}_n$
 can be seen as a direct
 consequence of the substitution principle \eqref{clt1}
 established in this paper.
Another major contribution of the paper is establish a new CLT for LSS for $\bS_n^0$
when the Gaussian-like moment conditions are not met.
In a related work, \cite{Pan2012}
 removes the Gaussian-like 4th-order moment condition,
but their assumptions of replacement,
made on both the population covariance matrices $\bSi_x$
and the Stieltjes transform of the LSD $F^{y,H}$,
are not easy to verify in applications.
 The new CLT of this paper removes the Gaussian-like 2nd-order and 4th-order moment condition restrictions and
 the given conditions are not only easy to satisfy but also are unremovable demonstrated by three counterexamples in Appendix.

We next address the same
problems for the class of Fisher matrices. From now on, for the
sample $\bbx_i$'s we will use the notations $\bSi_x=\bSi$ and
$\bS_x=\bS_n$. Consider another  sample
${\bby}_{1},\cdots,{\bby}_{m}$ of size $m$ from a $p$-dimensional
population $\bby$ with mean $\bnu$ and covariance matrix $\bSi_y$. The
corresponding  unbiased sample covariance matrix is
\begin{equation}
  \bS_y=\frac{1}{M}\sum\limits_{j=1}^m(\bby_j-\overline{\bby})(\bby_j-\overline{\bby})^*~,
  \label{Sy}
\end{equation}
where $\overline{\bby}=\frac1m\sum_j \bby_j$ is the sample mean
and $M:=m-1$ the adjusted sample size.
 The
so-called Fisher matrix $\bF:=\bS_x\bS_y^{-1}$ is a natural statistic
for the two-sample test of the hypothesis ``$\bSi_x=\bSi_y$''
that the populations have a same covariance matrix.
The CLT for LSS $\mu_{\bF}(g)$ of $\bF$
has been established  in \citet{Zheng2012} assuming
that both populations have zero means, i.e. $\bmu=\bnu=\mathbf{0}$
and standardized, i.e. $\bSi_x=\bSi_y=\bbI_p$. While keeping the
standardization assumption but dropping the condition
$\bmu=\bnu=\mathbf{0}$, we prove a similar substitution principle:
the CLT for LSS of $\bF:=\bS_x\bS_y^{-1}$ with arbitrary
population means and population distribution (provided that a
fourth-moment exists) is the same as the CLT in \cite{Zheng2012}
for populations with known means provided that one
substitutes the
  adjusted sample sizes $(N,M)=(n-1,m-1)$ for
  the sample sizes $(n,m)$
in the centering term of the CLT in
\cite{Zheng2012}. This second substitution principle can be viewed as
a consequence of the first substitution principle for sample
covariance matrices.

They have been other proposals in the literature for testing
hypotheses  about high-dimensional covariance matrices.
In particular, procedures are proposed in
\citet{CZZ10,LiChen12} using a family of well-chosen
$U$-statistics
and the asymptotic theory of these procedures does  not require
that $p/n$ tends to a positive limit.
In another perspective,  a minimax analysis for the one-sample
identity test  $\bSi_x=\bbI_p$ has been recently proposed in
\citet{CaiMa12}.
All these proposals are however not directly linked to the
substitutions principles discussed in this paper since
they do  not rely on LSS $\mu_{\bS_n}(g)$  or
$\mu_{\bF}(g)$  studied in this paper.

The  main results of the paper, the two substitution principles
and  the new CLT
are presented in
Sections~\ref{sec:one-sample} and \ref{sec:two-sample}. To
demonstrate the importance of these
principles, we develop in Section~\ref{testing} new
procedures
for hypothesis testing
about high-dimensional
covariance matrices
extending previous results to cover general
Gaussian or non-Gaussian populations with unknown populations means.
Technical proofs are relegated to Section
\ref{proofs}.

\section{Substitution principle for the unbiased sample covariance matrix
  $\bS_x$}

\label{sec:one-sample}

Before introducing the first substitution principle,
we give a new CLT of LSS for non-centered sample covariance matrix
whether the Gaussian-like moment conditions exist or don't exist.

\begin{description}
  \item[{\rm Assumption (a)}]
    Samples are $\{\bbx_j=\bmu+\bm{\Gamma}{\bf X}_j, j=1,\ldots,n\}$
    where ${\bf X}_j=(X_{1j},\ldots,X_{pj})'$.
  For each $p$, $\{X_{ij}, i\leq p,j\leq n\}$ are independent random
  variables with common moments $EX_{ij}=0$, $E|X_{ij}|^2=1$,
  $\beta_x=E|X_{ij}|^4-|EX_{11}^2|^2-2$, $\alpha_x=|EX_{11}^2|^2$ and satisfying the
  following Lindeberg condition:
  \begin{equation*}
    \frac{1}{np}\sum_{j=1}^p \sum_{k=1}^{n}
    E\{|X_{jk}|^4  \mathbf{1}_{\{|X_{jk}|\geq \eta \sqrt{n}\}}\}
    \rightarrow 0,
    \qquad \mbox{for any fixed } \eta>0
    \label{eq21}
  \end{equation*}
   where $\kappa=1$ is for complex $\{X_{ij}\}$ and $\kappa=2$ is for real $\{X_{ij}\}$.
  \item[{\rm Assumption (b)}]
    The dimension-to-sample  ratio
    $y_{n}=p/n$ tends to a positive $y>0$ as
    $n,~p\to\infty$.
  \item[{\rm Assumption (c)}]
    The sequence of $(\bSi_x=\bm{\Gamma}\bm{\Gamma}^*)_{p\ge 1}$ is bounded in spectral norm and
    the ESD $H_p:=\mu_{\bSi_x}$ of $\bSi_x$ converges weakly to
    a LSD  $H$ as $p\to\infty$.
    \item[{\rm Assumption (d1)}]
    ${\bm\Gamma}$ is real or complex $X_{ij}$ satisfies $\alpha_x=0$.
     \item[{\rm Assumption (d2)}]
     ${\bm \Gamma}^*{\bm\Gamma}$ is diagonal or $\beta_x=0$.
\end{description}
In fact, Assumption (d1) is for the 2nd-order moment condition of $X_{ij}$ and  Assumption (d2) is for the 4th-order moment condition of $X_{ij}$.
Assumption (d2) can be interpreted as follows: suppose the singular decomposition of $\bm{\Gamma}$ is $\bm{\Gamma} ={\bf U}^{*}{\bf L}^{1/2}
{\bf V}$, then $\bm{\Sigma} =\bm{\Gamma}\bm{\Gamma}^{*}={\bf U}^{*}{\bf L}{\bf U}$
where ${\bf L}$ is the diagonal matrix formed by eigenvalues and ${\bf U}^{*}$ by eigenvectors of $\bm{\Sigma}$. Then we can see that $\bm{\Gamma}^{*}\bm{\Gamma}={\bf V}^*{\bf L}{\bf V}$ is diagonal if the the unitary matrix $\bf V$ is an identity. It is the case especially when ${\bf VY}$ has the same distribution as $\bf Y$.
So it shows that Assumption (d2) is easy to satisfy.
Write $\bbx_j =\bmu+\bbx_j^0 $ with   $\bbx_j^0=\mathbf{\Gamma} {\bf X}_j$
and define the corresponding non-centered
sample covariance matrix as
\begin{equation}\label{Sx0}
  \bS_x^0 :=\frac1{n} \sum_{j=1}^n \bbx_j^0  {\bbx_j^0}^*~.
\end{equation}
Under Assumptions (a)-(b)-(c), it is well-known that both
the unbiased sample covariance matrix $\bS_x$ and non-centered sample covariance matrix $\bS_x^0$ have the same LSD $F^{y,H}$, namely
the  Mar\v{c}enko-Pastur distribution of index $(y,H)$.
We recall some useful facts about these distributions
(see \cite{BSbook} for details).
The LSD  has  support
\begin{equation}
  \label{eq:support}
  [a,b]=[(1-\sqrt{y})^2 I_{(0<y<1)}\lim\inf\limits_{n}\lambda_{\min}^{\bSi_x},~
  (1+\sqrt{y})^2\lim\sup\limits_{n}\lambda_{\max}^{\bSi_x}],
\end{equation}
where it has a density function. Moreover,
$F^{y,H}$
has a point mass $1-1/y$ at the origin when  $y>1$.
Define
$\underline{m}_{y}$ to be the Stieltjes transform of the companion
LSD $\underline F^{y,H} = (1-y)\delta_0 + y  F^{y,H}$. Then
$\underline{m}_{y}$ is the unique solution
in $\mathbb{C}^+=\{z:\Im(z)>0\}$ of the equation
\begin{equation}\label{Steiltjes}
  z=-\frac{1}{\underline{m}_{y}(z)}+y\int\frac{tdH(t)}{1+t\underline{m}_{y}(z)},~
  z\in \mathbb{C}^+=\{z:\Im(z)>0\}.
\end{equation}
Notice that when  a finite-horizon proxy
$F^{y_n,H_p}$ is substituted for  the LSD $F^{y,H}$,
these  properties and relationships hold  with
the parameters  $(y,H)$
replaced by $(y_n,H_p)$.

For Gaussian-like
moment conditions $\beta_x=0$ or $\alpha_x=0$ for complex population,
the CLT \eqref{clt0} for LSS of the non-centered sample covariance matrix $\bS_x^0$ has
been established first in \citet{BS04} where the explicit
limiting mean and covariance functions are given.
However, this result has a limitation in that it requires
Gaussian-like moment conditions, i.e., $\beta_x=0$ $\alpha_x=0$ for complex population.
There have been many efforts in the literature
for removing this restriction, see
\citet{LytPas09}  and \citet{PanZhou08}.
The CLT in \cite{PanZhou08}
removes the Gaussian-like 4th-order condition
$\beta_x=0$.
However, their assumptions of replacement,
made on both the population covariance matrices $\bSi_x$
and the Stieltjes transform of the LSD $F^{y,H}$,
are not easy to verify in applications.
Moreover, it is of practical importance to remove the Gaussian-like 2nd condition
but rare literature mentioned it.
In this section, we propose a new CLT under Assumptions (d2) and/or  (d1) without assuming
these Gaussian-like moment conditions made in \cite{BS04}.
Three counterexamples are provided in Appendix
to show that these assumptions (d1) and/or (d2)
can't be removed for a general CLT for LSS of the sample covariance matrix
$\bS_x^0$.

\begin{thm}\label{th:newCLT}
  Assume that either Assumptions (a)-(b)-(c)-(d1)-(d2) hold.
  Let
  $f_1,\ldots,f_k$  be  functions analytic on an open domain  of  the
  complex plan enclosing  the support of the LSD $F^{y,H}$ and
  define
  \begin{equation}\label{Yp}
    X_p(f_\ell) = p\left\{ \mu_{\bS_x^0}(f_\ell)-pF^{y_n,H_p}(f_\ell)  \right\}
    = \sum_{i=1}^p f_\ell(\lambda_j^0) - pF^{y_n,H_p}(f_\ell) ~,
  \end{equation}
  where $\{\lambda_j^0\}$ are eigenvalues of $\bS_x^0$. Then
  the random vector
  $(X_p(f_1),\ldots,X_p(f_k))$
  converges to a $k$-dimensional Gaussian
  random vector $(X_{f_1},\ldots,X_{f_k})$ with mean function
 \begin{eqnarray}
  \rE X_f&=&-\frac{1}{2\pi i}\oint\limits_{\cal C}
  f(z)\frac{\alpha_xy\int\frac{\underline{m}_{y}^3(z)t^2}{(1+t\underline{m}_{y}(z))^{3}}dH(t)}
  {\left(1-y\int\frac{\underline{m}_{y}^2(z)t^2}{(1+t\underline{m}_{y}(z))^{2}}dH(t)\right)\left(1-\alpha_xy\int\frac{\underline{m}_{y}^2(z)t^2}{(1+t\underline{m}_{y}(z))^{2}}dH(t)\right)}dz\nonumber\\
  &&-\frac{\beta_x}{2\pi i}\cdot\oint\limits_{\cal C}\left[f(z)\cdot\frac{y\underline{m}_{y}^3(z)\int\frac{t^2}
      {(\underline m_{y}(z)t+1)^3}dH(t)}{1-y\int\frac{\underline{m}_{y}^2(z)t^2dH(t)}
      {(1+t\underline{m}_{y}(z))^2}}\right]dz,\nonumber
\end{eqnarray}
and variance-covariance function
\begin{eqnarray}
  &&{\rm Cov}(X_f,
  X_g)\\
  &=&-\frac{1}{4\pi^2}\oint\limits_{{\cal
C}_1}\oint\limits_{{\cal
C}_2}\frac{f(z_1)g(z_2)}{(\underline{m}_{y}(z_1)-\underline{m}_{y}(z_2))^2}
  \frac{d}{dz_1}\underline{m}_{y}(z_1)\frac{d}{dz_2}\underline{m}_{y}(z_2) dz_1dz_2\label{covA3}\\
  &&-\frac{y\beta_x}{4\pi^2}\oint\limits_{{\cal
C}_1}\oint\limits_{{\cal
C}_2}\left[\int\frac{f_1(z_1)}{(\underline m_{y}(z_1)t+1)^2}
    \frac{f_2(z_2)}{(\underline m_{y}(z_1)t+1)^2}dH(t)\right]dz_1dz_2\nonumber\\
  &&-\frac{1}{4\pi^2}\oint\limits_{{\cal
C}_1}\oint\limits_{{\cal
C}_2} f_1(z_1)f_2(z_2)\left[\frac{d^2}{d z_1d z_2}\log(1-a(z_1,z_2))\right]dz_1dz_2
\end{eqnarray}
where ${\cal C}$, ${\cal C}_1$ and ${\cal C}_2$ are
closed contours in the complex plan enclosing  the support of
the LSD $F^{y,H}$, and ${\cal C}_1$ and ${\cal C}_2$
being non-overlapping.
Finally the function $a(z_1,z_2)$ is
\[
  a(z_1,z_2)
  =
  \alpha_x\left(1+\frac{\underline{m}_{y}(z_1)\underline{m}_{y}(z_2)(z_1-z_2)}{\underline{m}_{y}(z_2)-\underline{m}_{y}(z_1)}\right).
\]
\end{thm}

The proof of this refinement is given in
Section~\ref{pf:newCLT}.
Moreover, as said above, new Assumptions $(d1)$
and $(d2)$ are used as a replacement and they will be proven to be
necessary by examples shown in the appendix of the paper.
The major advantage  of this CLT is that the fourth-order and second-order population moments
can be arbitrary instead of matching Gaussian-like population, that is,
the parameters $\beta_x$ and $\alpha_x$ may be nonzero.

When the Gaussian-like 2nd-order moment condition ($\kappa=2, \alpha_x=1$ for real $\{X_{ij}\}$ and $\kappa=1, \alpha_x=0$ for complex $\{X_{ij}\}$) holds,
it can be easily checked that
the previous limiting mean and
variance-covariance functions
reduce to
\begin{eqnarray}
   \rE X_{f_j}&=&-\frac{\kappa-1}{2\pi i}\oint\limits_{\cal C}
   f_j(z)\frac{y\int\limits\underline{m}_{y}^3(z)t^2(1+t\underline{m}_{y}(z))^{-3}dH(t)}
   {[1-y\int\underline{m}_{y}^2(z)t^2(1+t\underline{m}_{y}(z))^{-2}dH(t)]^2}dz\nonumber\\
      &&-\frac{\beta_x}{2\pi i}\cdot\oint\limits_{\cal C}
      \left[f_j(z)\cdot\frac{y\int\frac{t^2\underline{m}_{y}^3(z)}{(\underline m_{y}(z)t+1)^3}dH(t)}
      {1-y\int\frac{\underline{m}_{y}^2(z)t^2dH(t)}{(1+t\underline{m}_{y}(z))^2}}\right]dz,
\end{eqnarray}
and variance function
\begin{eqnarray}
&&{\rm Cov}(X_{f_j},X_{f_{\ell}}) \nonumber\\
&=&-\frac{\kappa}{4\pi^2}\oint\limits_{{\cal
C}_1}\oint\limits_{{\cal
C}_2}\frac{f_j(z_1)f_{\ell}(z_2)}{(\underline{m}_{y}(z_1)-\underline{m}_{y}(z_2))^2}
d\underline{m}_{y}(z_1)d\underline{m}_{y}(z_2) \ \\
             &&-\frac{y\beta_x}{4\pi^2}\oint\limits_{{\cal C}_1}
             \oint\limits_{{\cal C}_2}f_j(z_1)f_{\ell}(z_2)
             \left[\int\frac{t}{(\underline m_{y}(z_1)t+1)^2
             }\frac{t}{(\underline m_{y}(z_1)t+1)^2}dH(t)\right]d\underline{m}_{y}(z_1)d\underline{m}_{y}(z_2).\nonumber
\end{eqnarray}
In particular, under Gaussian-like 2nd-order and 4th-order moment conditions,
we recover the CLT~\eqref{clt0} of \cite{BS04}.

Coming to the unbiased sample covariance matrix $\bS_x$ with unknown population means,
as a second main result of the paper, we establish the following
substitution principle.
Recall that
$N=n-1$ denotes the adjusted sample size.

\begin{thm}\label{th:substitu1}
(One sample substitution principle) \quad Under the same
conditions as in Theorem \ref{th:newCLT}, define
\begin{equation}\label{Xp}
  Y_p(f_\ell) = p\left\{ \mu_{\bS_x}(f_\ell) - F^{y_{N},H_p}(f_\ell)\right\}=
    \sum_{i=1}^p f_\ell(\lambda_j) - pF^{y_{N},H_p}(f_\ell) ~,
\end{equation}
where $\{\lambda_j\}$ are the eigenvalues of the unbiased sample
covariance matrix $\bS_x$ and $N=n-1$. Then the random vector
$(Y_p(f_1),\ldots,Y_p(f_k))$ converges in distribution to the same
Gaussian vector $(X_{f_1},\ldots,X_{f_k})$ given in Theorem
\ref{th:newCLT}.
\end{thm}

The proof of Theorem~\ref{th:substitu1}  is postponed to Section~\ref{pf:substitu1}.

\section{Substitution principle for the two-sample Fisher matrix}
\label{sec:two-sample} In this section we investigate the effect
in the CLT for LSS of $\bF=\bS_x\bS_y^{-1}$ when the unbiased
covariance matrices $\bS_x$ and $\bS_y$ are used. The following
assumptions for the second sample $\bby_1,\ldots,\bby_m$ mimic
Assumptions (a)-(b)-(c) set for the first sample
$\bbx_1,\ldots,\bbx_n$.

\begin{description}

\item[{\rm Assumption (a')}] Samples are $\{\bby_j=\bnu+\bGa_y  {\bf Y}_j, j=1,\ldots,m\}$
where ${\bf Y}_j=(Y_{1j},\ldots,Y_{pj})$.
  For each $p$, the elements of the data matrix
    $\{Y_{ij}, i\leq p,j\leq m\}=\{Y_1,\ldots,Y_m\}$ are independent random
    variables with common moments $EY_{ij}=0$, $E|Y_{ij}|^2=1$,
    and $E|Y_{ij}|^4=\beta_y+2+|EY_{11}^2|^2$, especially $\E Y^2_{ij}=0$ in complex case and satisfying the
    following Lindeberg condition:
    \begin{equation*}
      \frac{1}{mp}\sum_{j=1}^p \sum_{k=1}^{m}
      E\{|Y_{jk}|^4  \mathbf{1}_{\{|Y_{jk}|\geq \eta \sqrt{m}\}}\}
      \rightarrow 0,
      \qquad \mbox{for any fixed } \eta>0.
      \label{eq24}
    \end{equation*}
  \item[{\rm Assumption (b')}]
    The dimension-to-sample  ratio
    $y_{m}=p/m \to y_2\in(0,1)$ as
    $m,~p\to\infty$.
  \item[{\rm Assumption (c')}]
    The sequence  $(\bSi_y)_{p\ge 1}$ where $\bSi_y=\bGa_y\bGa_y^*$ is bounded in spectral norm and
    the ESD $H_{2,p}$ of $\bSi_y$ converges to
    a LSD  $H_2$ as $p\to\infty$.
\end{description}

Regarding the distinction between real-valued and complex-valued
populations, a same indicator $\kappa$ is used for both
populations $\bbx$ and $\bby$ since
the mixed situation where one population is real-valued while the
other is complex-valued is rarely realistic in applications.

Consider first the non-centered and
sample covariance matrices
\begin{equation} \label{F0}
  \bS_X^0 = \frac1n \sum_{j=1}^n X_j X_j^*, \quad
  \bS_Y^0 = \frac1m \sum_{j=1}^m Y_j Y_j^*~,\quad
  \bF^0 = {\bS_X^0} {\bS_Y^0}^{-1}.
\end{equation}
Assume that  Assumptions (a)-(b)-(c) and (a')-(b')-(c') are
fulfilled.
In this section, if both populations are complex,
we assume that the second moments are null, i.e.
$\E X_{ij}^2=\E Y_{ij}^2=0$.
From now onward, for notation convenience,
the limiting ratio $y=\lim p/n$ of the $\bbx$-sample is denoted
by $y_1$.
It is well-known from random matrix theory
that the ESD of $\mu_{\bF^0}$ converges to a
LSD $G_{(y_1,y_2)}$ with compact support
(\citet{BaiYinKri87,Silv95}).  Moreover, let
 $f_1,\ldots,f_k$ be   analytic functions  on an
open set of  the complex plan enclosing  the support of
$G_{(y_1,y_2)}$. Consider linear spectral statistics
\begin{equation}\label{Zp}
  Z_p(f_\ell) = p\left\{ \mu_{\bF^0} ( f_\ell) -
    G_{(y_{n},y_m)}(f_\ell) \right\}~,
\end{equation}
where, similar to CLT's  for sample covariance matrices,
$G_{(y_{n},y_m)}$ is a finite-horizon proxy for
the LSD $G_{(y_1,y_2)}$ obtained by substituting  the
current  dimension-to-sample ratios
$(y_n,y_m)=(p/n,p/m)$ for their limits
$(y_1,y_2)=\lim ~(p/n,p/m)$.
Let $h=(y_1^2+y_2^2+y_1y_2)^{1/2}$.
Then the  CLT in \citet{Zheng2012} establishes that
the random vector
$(Z_p(f_1),\ldots,Z_p(f_k))$ converges to a $k$-dimensional
Gaussian vector $(Z_{f_1},\ldots,Z_{f_k})$
with mean function
\begin{eqnarray}
  &&\rE Z_{f_j}\nonumber\\
  &=&
  \lim_{r\downarrow 1}\frac{\kappa-1}{4\pi
    i}\ointctrclockwise\limits_{|\xi|=1}
  f_j\left(\frac{1+h^2+2h\Re(\xi)}{(1-y_2)^2}\right)
  \left[\frac{1}{\xi-r^{-1}}+\frac{1}{\xi+r^{-1}}-\frac{2}{\xi+\frac{y_2}{h}}\right]~~d\xi
\label{m1bis}\\
    && +\frac{\beta_xy_1(1-y_2)^2}{2\pi i\cdot
    h^2}\ointctrclockwise\limits_{|\xi|=1}
  f_j\left(\frac{1+h^2+2h\Re(\xi)}{(1-y_2)^2}\right)
  \frac{1}{(\xi+\frac{y_2}{h})^3}~~d\xi~,
  \nonumber\\
   && +\frac{\beta_y(1-y_2)}{4\pi
    i}\ointctrclockwise\limits_{|\xi|=1}
  f_j\left(\frac{1+h^2+2h\Re(\xi)}{(1-y_2)^2}\right)
  \frac{\xi^2-\frac{y_2}{h^2}}{(\xi+\frac{y_2}{h})^2}
  \left[\frac{1}{\xi-\frac{\sqrt{y_2}}{h}}
   +\frac{1}{\xi+\frac{\sqrt{y_2}}{h}}-\frac{2}{\xi+\frac{y_2}{h}}\right]d\xi,\nonumber
\end{eqnarray}
and covariance function
\begin{eqnarray}
  &&\cov(Z_{f_j},Z_{f_{\ell}})\nonumber\\
  &=&
    -\displaystyle{\lim_{r\downarrow 1}\frac{\kappa}{4\pi^2}\ointctrclockwise\limits_{|\xi_1|=1}
    \ointctrclockwise\limits_{|\xi_2|=1}
    \frac{f_j\left(\frac{1+h^2+2h\Re(\xi_1)}{(1-y_2)^2}\right)
      f_{\ell}\left(\frac{1+h^2+2h\Re(\xi_2)}{(1-y_2)^2}\right)}{(\xi_1-r\xi_2)^2}~d\xi_1d\xi_2}
  \label{cov1bis}\\
  &&  -\frac{(\beta_xy_1+\beta_yy_2)(1-y_2)^2}{4\pi^2
    h^2}\ointctrclockwise\limits_{|\xi_1|=1}
  \frac{f_j\left(\frac{1+h^2+2h\Re(\xi_1)}{(1-y_2)^2}\right)}
       {(\xi_1+\frac{y_2}{h})^2}d\xi_1
       \ointctrclockwise\limits_{|\xi_2|=1}\frac{f_{\ell}\left(\frac{1+h^2+2h\Re(\xi_2)}
         {(1-y_2)^2}\right)}{(\xi_2+\frac{y_2}{h})^2}d\xi_2.\nonumber
\end{eqnarray}

For the
Fisher matrix of interest $\bF=\bS_x\bS_y^{-1}$ from
populations with unknown population means and as the second main result of the paper,
we establish  the
following substitution principle under an additional condition of
equal covariance matrix.

\begin{thm}\label{th:substitu2}(Two-sample substitution principle)\quad
  Assume that the Assumptions (a)-(b)-(c) and
  (a')-(b')-(c') are fulfilled with $y_2\in(0,1)$
  and that $\bGa=\bGa_y$.
  Let $f_1,\ldots,f_k$ be  functions analytic  on an
  open domain  of  the complex plan enclosing
  the support of the LSD $G_{(y_1,y_2)}$ and define linear spectral
  statistics
  \begin{equation}\label{Zp}
    W_p(f_\ell) =
    p \left\{ \mu_{\bF} (f_\ell) - G_{(y_{N},y_{M})}(f_\ell) \right\}~,
  \end{equation}
  where $N=n-1$ and $M=m-1$ are the adjusted sample sizes,
  $y_N=p/N$ and $y_M=p/M$.
  Then the random vector  $(W_p(f_1),\ldots,W_p(f_k))$
  converges to the same  limiting  $k$-dimensional Gaussian vector
  $(Z_{f_1},\ldots,Z_{f_k})$ defined \cite{Zheng2012}
   with
  the mean and covariance functions \eqref{m1bis}-\eqref{cov1bis}.
\end{thm}
The proof of this theorem is given in Section~\ref{pf:substitu2}.

\section{Applications to hypothesis testing  on  large  covariance matrices}
\label{testing}

As explained in Introduction, this section is
devoted to illustrate the importance of the substitution principles
proposed in this paper.
We consider the problem of testing hypotheses about large covariance
matrices based on the unbiased sample
covariance matrices when population means are to be estimated.
In this manner, Sections \S\ref{test1} and
\S\ref{test2} generalize the main results of \citet{BJYZ09}
on the one-sample and two-sample likelihood ratio
tests on large covariance matrices.
The generalized  test procedures
apply for non-Gaussian populations with unknown population means. To our
best knowledge,
few procedures exist for  such testing problems
on large sample covariance matrices, two exceptions being
\citet{CZZ10},
\citet{LiChen12}, see also  \citet{CaiMa12} on a  minimax
study for the identity test.

\subsection{Testing the hypothesis that $\bSi$ is equal to a given  matrix}
\label{test1}

Let as in Introduction $\bbx_1,\ldots,\bbx_n$ be a sample from a
$p$-dimensional population with mean $\bmu$ and  covariance matrix
$\bSi_x$. Consider first a one-sample test for the hypothesis
$H_0:~\bSi_x={\bbI_p}$ that a $p$-dimensional covariance matrix
$\bSi_x$ equals the identity matrix. The {\em corrected likelihood
ratio} test in \citet{BJYZ09} is  developed by assuming that the
population is Gaussian and $\bmu=0$ (or equivalently, $\bmu$ is given). The
test statistic equals
\begin{equation}
   {L^0}=\mbox{tr~} \textbf{S}_x^0 -
  \log|\textbf{S}_x^0|-p,   \label{stat10}
\end{equation}
where $\bS_x^0$ is  the non-centered sample covariance matrix given in
\eqref{Sx0}. The following theorem is established in
\cite{BJYZ09}.
\begin{prop} (Theorem 3.1 of \cite{BJYZ09})\label{Th3.1}
  Assume that the population is real Gaussian with mean $\bmu=0$ and
  covariance matrix $\bSi_x$, and the dimension $p$ and the sample
  size tend to infinity such that $y_n:=p/n\to y\in (0,1)$.
  Then
  under $H_0$,
  \begin{equation}
    \upsilon_n(g)^{-\frac{1}{2}}\left[   L^0-p \cdot
      F^{y_n}(g)- m_n(g)\right] \Rightarrow N \left( 0,
    1\right),
    \label{singlsta0}
  \end{equation}
  where $F^{y_n}$ is the Mar\v{c}enko-Pastur law of  index  $ y_n$,
  $g(x)=x-\log x-1$ and
  \begin{eqnarray*}
     F^{y_n}(g) & = & 1-\frac{y_n-1}{y_n}\log{(1-y_n)}~, \\
     m_n(g)& =& -\frac{\log{(1-y_n)}}{2} ~, \\
     \upsilon_n(g)& =    & -2\displaystyle\log{(1-y_n)}-2y_n~.
  \end{eqnarray*}
\end{prop}
At asymptotic significance level $\alpha$, the test will reject
the null hypothesis if the statistic in \eqref{singlsta0} exceeds
$z_\alpha$, the upper $\alpha$\% quantile of the standard Gaussian
distribution. The test has been proved to have good powers against
the inflation of the dimension $p$. To extend this result to
general populations with unknown population mean vector, we start
by assuming that the population $\bbx$ fulfills Assumption
(a)-(b)-(c) of Section~\ref{sec:one-sample}. The corrected
likelihood ratio test statistic (CLRT) is defined to be
\begin{equation}
   {L^*}=\mbox{tr~} \textbf{S}_x -
  \log|\textbf{S}_x|-p,   \label{stat1}
\end{equation}
where $\bS_x$ is  the unbiased  sample covariance matrix given  in
\eqref{Sx}.

\begin{thm}\label{th:onesample}
Assume that the population $\bbx$ fulfills Assumptions (a)-(b)-(c)
where $y_n:=p/n\to y \in (0,1)$. Then under the null hypothesis
$H_0:~\bSi_x={\bbI}_p$, for the unbiased sample covariance matrix
$\bS_x$ in \eqref{Sx} and the LRT statistic $L^* $ in
\eqref{stat1}, we have,
\begin{equation}
  \upsilon^*_N(g)^{-\frac{1}{2}}\left[   L^*-p \cdot
    F^{y_{N}}(g)- m^*_N(g)\right] \Rightarrow N \left( 0,
  1\right)
  \label{singlsta}
\end{equation}
where
\begin{eqnarray*}
  m^*_N(g)& =& (\kappa-1) m_N(g)  + \frac{\beta_x}{2}y_N~, \\
  \upsilon^*_N(g)& =    & \frac{\kappa}{2} v_N(g) ~,
\end{eqnarray*}
and the function $g$, the  values $F^{y_{N}} (g)$, $m_N(g)$ and
$v_N(g)$ are the same as in Proposition~\ref{Th3.1} (notice
however the substitution of $N$ for $n$ in these quantities).
\end{thm}
Let us explain how this result extends considerably the previous
Proposition~\ref{Th3.1} proposed in  \citet{BJYZ09}. For real
Gaussian observations, we have $\kappa=2$ and $\beta_x=0$, then
$m^*_N(g) =m_N(g)$ and $\upsilon_N^*(g)= \upsilon_N(g)$, so that
the new  CLT gives an extension of
Proposition~\ref{Th3.1} to Gaussian populations.
If the
observations are complex Gaussian, $\kappa=1$ and $\beta_x=0$, we
have $m^*_N(g)=0$ and the variance
$\upsilon^*_N(g)=\frac12\upsilon_N(g)$, which is  half the
variance for the real Gaussian case. For general non-Gaussian
and non-centered populations,
the new CLT provides a novel
procedure for the one-sample test on large
covariance matrix.
In this case,
the variance $\upsilon^*_N(g) $ stays the same as
for Gaussian observations, but there is an additional term
$\frac12{\beta_x}y_N$ in the asymptotic mean.

We conclude the section by reporting a small Monte-Carlo
experiment that demonstrates the importance of the sample size
substitution proposed in Theorem \ref{th:onesample}. We simulate a
standard Gaussian population $\mathbf{x} \sim N({\bf 0}_p, {\bf
I}_p)$ but we don't assume to know anything about the mean and the
covariance matrix so that the test will be based on the statistic
$L^*$ of \eqref{stat1}. Simulation results are listed in
Table~\ref{tab:sph1}.

{\small
\begin{center}
  \begin{table}[hptb]
    \caption{Effects of the sample size substitution
      for the corrected
      one-sample LRT \label{tab:sph1} with unbiased covariance matrix.
      Standard
      normal population with 10000 independent replications.}
    \renewcommand{\arraystretch}{1.1}   \doublerulesep 2pt \tabcolsep
    0.04in
    \begin{tabular}{l|ccc}
      \hline
      &$(pF^{y_{N}}(g)+m^*_N(g), v^*_N(g))$& $(pF^{y_n}(g)+m_n^*(g), v_n^*(g))$ & Empirical mean and \\
      & &  &variance of $ {L}^*$\\   \\ \hline
      &\multicolumn{3}{c}{$p/n=0.5$}\\ \\
      $(p,~n)$=(25, 50)  &(8.226, 0.407) &(8.017, 0.386)&(8.234, 0.452)\\
      \\
      $(p,~n)$=(50, 100) &(15.889, 0.396)  &(15.689, 0.386)  &(15.886, 0.405)\\
      \\
      $(p,~n)$=(100, 200)&(31.228, 0.391)  &(31.031, 0.386)  &(31.231, 0.410)\\
      \\
     $(p,~n)$=(150, 300)&(46.570, 0.390)  &(46.374, 0.386)  &(46.569, 0.404)\\
      \\ \hline
      &\multicolumn{3}{c}{$p/n=0.8$}\\ \\
      $(p,~n)$=(32, 40) &(20.835, 1.794)&(19.929, 1.618)&(20.895, 2.158)\\
     \\
     $(p,~n)$=(64, 80) &(39.909, 1.702) &(39.053, 1.618) &(39.931, 1.851)\\
      \\
      $(p,~n)$=(96, 120)&(59.018, 1.673)&(58.178, 1.618)&(59.051, 1.739)\\
      \\
      $(p,~n)$=(128, 160)&(78.135, 1.659)&(77.302, 1.618) &(78.132, 1.714)\\
      \\
      \hline
    \end{tabular}
  \end{table}
\end{center}
}

 For the distribution of the CLRT statistic $L^*$, the
experiment shows that the formula for its  asymptotic mean and
variance with adjusted dimension-to-sample ratio $y_{N}=p/ N $
always outperforms the formula without the adjustment using
$y_n=p/n$. The difference is quite significant for $p/n=0.8$. This
is an interesting improvement since when $p/n$ is getting close to
1, the sample covariance matrix has more small eigenvalues near 0
and the presence of the logarithm function in the LRT statistic
makes it  more sensible with a larger variance. So a more accurate
approximation for its asymptotic distribution is particularly
valuable in such situations.

\subsection{Testing the equality of two large covariance matrices}\label{test2}
The second test problem we consider is about the  equality between
two large covariance matrices. As in
Section~\ref{sec:two-sample}, let $\bbx_1,\ldots,\bbx_n$ and
$\bby_1,\ldots,\bby_m$ be samples from two $p$-dimensional
populations with mean and covariance matrix $(\bmu,\bSi_x)$ and
$(\bnu,\bSi_y)$, respectively. To test the hypothesis $  H_0:~
\bSi_x=\bSi_y$, a  {\em corrected likelihood ratio} test is
developed in \citet{BJYZ09} by assuming that both  populations are
Gaussian and $\bmu=\bnu=0$ (or equivalently, they  are  given). Under the
null hypothesis and because of the Gaussian assumption, one can
assume without loss of generality that $ \bSi_x=\bSi_y=\bbI_p$.
Therefore the sample covariance matrices $\bS_X^0$ and $ \bS_Y^0 $
are as defined in \eqref{F0} and the normalized Fisher matrix is
$\bF^0 = {\bS_X^0} {\bS_Y^0}^{-1}$. The LRT statistics is
\begin{equation}
  {T^0}
  =\frac{\left|{\bS_X^0}\right|^{\frac{n}{2}}\cdot
    \left|\bS_Y^0\right|^{\frac{m}{2}}}
  {\left|c_1{\bS_X^0}+c_2\bS_Y^0\right|^{\frac{n+m}{2}}},\label{S1S2CD}
\end{equation}
where $c_1= n/(n+m)$ and  $c_2= m/(n+m)$. Recall the ratios
$y_{n}:=\frac{p}{n}$, $y_{m}:=\frac{p}{m}$ and set
\[
h_n = (y_{n}+y_{m}- y_{n}y_{m})^{1/2}~.
\]
The following result is established in \cite{BJYZ09}.

\begin{prop} (Theorem 4.1 of \cite{BJYZ09})\label{T4.1}
  Assume that  both populations are  real Gaussian with respective
  mean 0
  and
  covariance matrices $\bSi_k$, $k=1,2$,  and that
  $p\wedge n\wedge m \to \infty $ such that
  $y_{n}\rightarrow y_1>0$,
  $y_{m}\rightarrow y_{2}\in (0, 1).$
  Then  under $H_0$,
  \begin{equation}
    \upsilon_{n,m}(f)^{-\frac{1}{2}}\left[
      -\displaystyle\frac{2\log T^0}{n}-p \cdot G_{y_{n},y_{m}}(f)-
     a_{n,m}(f)\right] \Rightarrow N \left( 0, 1\right)\label{LST}
  \end{equation}
  where
  \begin{eqnarray}
    f(x)&=&\log(y_{n}+y_{m}x)-\frac{y_{m}}{y_{n}
      +y_{m}}\log x-\log(y_{n}+y_{m})~,  \label{fn1n2}\\
    {G_{y_{n}, y_{m}}(f)} &=&
    \frac{h_n^2}{y_{n}y_{m}}  \log \frac {y_{n}+y_{m}}  { h_n^2 }
    +{\frac{y_{n}(1-y_{m})}{y_{m}(y_{n}+y_{m})}\log{(1-y_{m})}}\nonumber\\
    &&+{\frac{y_{m}(1-y_{n})}{y_{n}(y_{n}+y_{m})}\log{(1-y_{n})}}, \label{Fyn1yn2}
   \\
   a_{n,m}(f)&=&
    \frac{1}{2}\left[\log\left(\frac{h_n^2}{y_{n}+y_{m}}\right)
      -\frac{y_{n}}{y_{n}+y_{m}}\log(1-y_{m}) \right.\nonumber\\
      &&
      \left.
      -\frac{y_{m}}{y_{n}+y_{m}}\log(1-y_{n})\right]~,
    \label{testE}\\
    \upsilon_{n,m}(f)&=&-\frac{2y_{m}^2}{(y_{n}+y_{m})^2}\log(1-y_{n})-\frac{2y_{n}^2}{(y_{n}+y_{m})^2}\log(1-y_{m})\nonumber\\
    &&+2\log\frac{h_n^2}{y_{n}+y_{m}}\label{testVar}.
  \end{eqnarray}
\end{prop}
Again, a corrected LRT is obtained based on this limiting
distribution and has been proved to have good powers for large
dimensions $p$. To extend this result to general non-Gaussian populations with unknown population means, we
start by assuming that the
population $\bbx$ fulfills Assumption (a)-(b)-(c) of
Section~\ref{sec:one-sample} and the
 population $\bby$ fulfills Assumption (a')-(b')-(c') of
Section~\ref{sec:two-sample} The corrected likelihood ratio test
statistic (CLRT) is defined to be
\begin{equation}
  {T^*}=\frac{\left|{\bS_x}\right|^{\frac{n}{2}}\cdot
    \left|\bS_y\right|^{\frac{m}{2}}}
  {\left|c_1{\bS_x}+c_2\bS_y\right|^{\frac{n+m}{2}}},\label{stat2}
\end{equation}
with the constants $c_k $ defined previously. Here the unbiased
sample covariance matrices $ \bS_x$  and $ \bS_y$ are defined in
\eqref{Sx} and \eqref{Sy}, respectively.

\begin{thm} \label{th:twosample}
  \quad
  Assume that the populations $\bbx$ and $\bby$ satisfy Assumptions
  (a)-(b)-(c)
  and (a')-(b')-(c'), respectively.
  Then
  under the null hypothesis $H_0:~ \bSi_x=\bSi_y$,
  \begin{equation}
    \upsilon^*_{N,M}(f)^{-\frac{1}{2}}\left[
      -\displaystyle\frac{2\log T^*}{n}-p \cdot G_{y_{N},y_{M}}(f)-
      a^*_{N,M}(f)\right] \Rightarrow N \left( 0, 1\right)~,
 \end{equation}
  where
  \begin{eqnarray*}
    a_{N,M}^*(f) &=&  (\kappa-1) a_{N,M}(f) +\frac{y_{N} y_{M}}{2(y_{N}+y_{M})^2}
    (\beta_x y_{N} +\beta_yy_{M})~,\\
    \upsilon_{N,M}^*(f) & = & \frac{\kappa}2 \upsilon_{N,M}(f) ~,
  \end{eqnarray*}
  where the function
  $f(x)$, and the values
  $G_{y_{N}, y_{M}}(f)$,  $a_{N,M}(f)$  and $ \upsilon_{N,M}(f)$ are the same
  as in Proposition~\ref{T4.1}
  (notice however the
  substitution of
   $(N,M)=(n-1,m-1)$
   for $(n,m)$ in these formula).
\end{thm}

Again it is interesting to compare this CLT to the previous one in
Proposition~\ref{T4.1}. When both populations are real Gaussian,
$\kappa=2$ and $\beta_x=\beta_y=0$, we have
$a_{N,M}^*(f)=a_{N,M}(f)$ and $ \upsilon_{N,M}^*(f)=
\upsilon_{N,M}(f)$, the new CLT
is an extension of
Proposition~\ref{T4.1} to Gaussian populations.
When there are both complex Gaussian,
$\kappa=1$ and $\beta_x=\beta_y=0$,   $a_{N,M}^*(f)=0$ and the
variance $ \upsilon_{N,M}^*(f)$ is reduced by half. For general
non-Gaussian populations with unknown population means, there will be always a shift in the mean,
but the variance again remains {\em unchanged} compared to the
Gaussian situation.
  In summary, the substitution principle allows a full generalization of
  the corrected likelihood ratio two-sample test for
  large covariance matrices
  from  non-Gaussian populations with unknown population means.

We conclude the section by reporting a small Monte-Carlo
experiment to examine the effect of the sample size substitution
proposed in Theorem~\ref{th:twosample}. We adopt standard Gaussian
population for both populations $\mathbf{x},~ \bby \sim N({\bf
0}_p, {\bf I}_p)$ but we don't assume to know anything about these
parameters so that the test will be based on the statistic $T^*$
of \eqref{stat2}. Simulation results are listed in
Table~\ref{tab:sph2}.

{\small
\begin{center}
  \begin{table}[hptb]
   \caption{Effects of the sample size substitution
      for the corrected
      two-sample LRT \label{tab:sph2} with unbiased covariance
      matrices. Standard
      normal populations with  10000 independent replications.}
   \renewcommand{\arraystretch}{1.0}   \doublerulesep 2pt \tabcolsep   0.04in
    \begin{tabular}{l|ccc}
      \hline
      &$(pF^{y_{N}, y_{M}}(f)+a_{N,M}^*(f),$  & $(pF^{y_{n}, y_{m}}(f)+a^*_{n,m}(f),$  & Empirical mean and \\
      & $v^*_{N,M}(f))$ & $v^*_{n,m}(f))$ &variance of $T^*$\\   \\ \hline
      &\multicolumn{3}{c}{$p/n=0.5$}\\ \\
      $(p,~n)=(20,40)$  &(3.731, 0.127) &(3.601, 0.118) &(3.729, 0.134)\\
      \\
      $(p,~n)=(50,100)$  &(8.820, 0.121) &(8.698, 0.118)&(8.819, 0.122)\\
      \\
      $(p,~n)=(80,160)$ &  (13.916, 0.120)    &    (13.795, 0.118)     &(13.918, 0.125)\\
      \\ \hline
      &\multicolumn{3}{c}{$p/n=0.8$}\\ \\
      $(p,~n)=(20,25)$  &(8.551, 0.714) &(7.827, 0.588)&(8.520, 0.776)\\
      \\
      $(p,~n)=(60,75)$  &(23.011, 0.625) &(22.382, 0.588)&(23.012, 0.661)\\
      \\
      $(p,~n)=(100,125)$ &(37.550, 0.610) &(36.937, 0.588)&(37.552, 0.616)\\
      \\
      \hline
    \end{tabular}
  \end{table}
\end{center}
}

For the distribution of the CLRT statistic $T^*$, the limiting
parameters with adjusted dimension-to-sample ratios $y_{N}$ and
$y_{M}$ are much more accurate than using the original  ones
$y_{n}$ and $y_{m}$.

\section{Proofs} \label{proofs}
Some of the proofs below use several technical lemmas which are
collected and proved in Section~\ref{sec:lemmas}.
\subsection{Proof of  Theorem~\protect\ref{th:onesample}}
\label{pf:onesample} Under the null hypothesis $\bSi_x={\bf I}_p$
and by the substitution principle of Theorem~\ref{th:substitu1},
it is enough to consider the sample covariance matrix
\[ \bS_N^0=
\frac1N \sum_{i=1}^N {\bf X}_i{\bf X}_i ^* ~,
\]
where the ${\bf X}_i$'s have i.i.d. (0,1) components. Applying the
formula in Proposition~\ref{thm201302}, we only need to evaluate
the mean and variance parameter in
Eqs.~\eqref{eq:E}-\eqref{eq:cov} with the function $g(x)=x-\log
x-1$, i.e.
\[ m_N^*(g) = (\kappa-1) I_1(g)+\beta I_2(g) ~,
\]
and
\[ \upsilon^*_N(g) = \kappa J_1(g,g)+\beta J_2(g,g)~.\]
Note that the forms $\{I_\ell\}$ and $\{J_\ell \}$ are linear and
bi-linear, respectively, and  null on constants. Using their
values on the functions $x$ and $\log x$ calculated in Eq. (4.1)
to Eq. (4.10) in \citet{WY13}, we readily find the claimed formula
for   $m_N^*(g)$ and $\upsilon_N^*(g)$.

\subsection{Proof of   Theorem~\protect\ref{th:twosample}}
\label{pf:twosample} Under the null hypothesis, according to
\citet{BJYZ09}, the likelihood ratio statistic $-2n^{-1}\log T^*$
is a LSS of a Fisher matrix. Moreover, by the substitution
principle of Theorem~\ref{th:substitu2}, it is enough to consider
a Fisher matrix with dimension-to-sample ratios $y_N=p/N$ and
$y_M=p/M$ (instead of $y_n$ and $y_m$). We thus use the CLT of
\citet{Zheng2012} with these ratios and the test function $f$
defined in \eqref{fn1n2}, namely,
\[
f(x)=\log(y_{N}+y_{M}x)-\frac{y_{M}}{y_{N}+y_{M}}\log(x)-\log(y_{N}+y_{M})~.
\]
  Define
  $$
  f_1(x)=\log(y_{N}+y_{M}x),~f_2(x)=\log(x)~,
  $$
  so that $f=f_1- \frac{y_{M}}{y_{N}+y_{M}}  f_2-\log (y_{N}+y_{M})$. The asymptotic
  mean  $E(X_{f_k})$ and the variance-covariance
  functions $\cov(X_{f_k}, X_{f_\ell})$, $k,\ell=1,2$ are found using
  the calculations done
  in  Example 4.1
  of \citet{Zheng2012}
  with the following values of the parameters
  $c,~d,~c'$ and $d'$:
  \[ c'=1, \quad d'=h_n,\quad c=h_n, \quad d=y_{M}~.\]
  That is, the mean function is
 \begin{eqnarray*}
    EX_{f_1}& = & \frac{\kappa-1}2\log\left(\frac{(c^2-d^2)h_n^2}{(ch_n-y_{M}d)^2}\right)-\frac{\beta_xy_{N}(1-y_{M})^2d^2}
    {2(ch_n-dy_{M})^2} \\
    &&
    +\frac{\beta_y(1-y_{M})}{2}\left[\frac{2dy_{M}}{ch_n-dy_{M}}+\frac{d^2(y_{M}^2-y_{M})}{(ch_n-
        dy_{M})^2}\right],
    \\
    &=& \frac{\kappa-1}2\log\frac{h_n^2}{(y_{N}+y_{M})(1-y_{M})}
    -\frac{\beta_x}{2}\frac{  y_{N} y_{M}^2    }{(y_{N}+y_{M})^2}
    +\frac{\beta_y}{2}\frac{  y_{M}^2 (2y_{N}+y_{M})
    }{(y_{N}+y_{M})^2}~,
  \end{eqnarray*}
  and
  \begin{eqnarray*}
    EX_{f_2}& =& \frac{\kappa-1}2\log\left(\frac{((c')^2-(d')^2)h_n^2}{(c'h_n-y_{M}d')^2}\right)-\frac{\beta_xy_{N}(1-y_{M})^2(d')^2}
    {2(c'h_n-d'y_{M})^2}
    \\
    &&
    +\frac{\beta_y(1-y_{M})}{2}\left[\frac{2d'y_{M}}{c'h_n-d'y_{M}}
      +\frac{(d')^2(y_{M}^2-y_{M})}{(c'h_n-d'y_{M})^2}\right] \\
    &=& \frac{\kappa-1}2\log\frac{1-y_{N}} {1-y_{M}}
    -\frac12 {\beta_x}    y_{N}
    +\frac12 {\beta_y}      y_{M}  ~.
 \end{eqnarray*}
  And the variance function is
  \begin{eqnarray*}
    \var(X_{f_1})
    & =& \kappa\log\left(\frac{c^2}{c^2-d^2}\right)+
    \frac{(\beta_xy_{N}+\beta_yy_{M})(1-y_{M})^2d^2}         {(ch-dy_{M})^2}\\
    &=& \kappa\log\frac{h_n^2}{   (y_{N}+y_{M})(1-y_{M})    }
    + (\beta_xy_{N}+\beta_yy_{M}) \frac{y_{M}^2} {( y_{N}+y_{M} )^2}~, \\
    \var(X_{f_2}) &= &\kappa\log\left(\frac{(c')^2}{(c')^2-(d')^2}\right)+
    \frac{(\beta_xy_{N}+\beta_yy_{M})(1-y_{M})^2(d')^2}         {(c'h-d'y_{M})^2}\\
    &=& \kappa\log\frac{1}{   (1-y_{N})    (1-y_{M}) }+
    (\beta_xy_{N}+\beta_yy_{M})  ~,\\
    \cov(X_{f_1},X_{f_2})&=&\kappa\log\left(\frac{cc'}{cc'-dd'}\right)+
    \frac{(\beta_xy_{N}+\beta_yy_{M})(1-y_{M})^2dd'}         {(ch_n-dy_{M})(c'h_n-d'y_{M})}\\
    &=& \kappa\log\frac{1}{   1-y_{N}  }
    + (\beta_xy_{N}+\beta_yy_{M}) \frac{y_{M}} {y_{N}+y_{M} }~.
  \end{eqnarray*}
  As  by definition of $f$,
  \[
  a_{N,M}^*(f)=EX_{f_1}-\frac{y_{N}}{y_{N}+y_{M}}\cdot EX_{f_2}-\log(y_{N}+y_{M}),
  \]
  and
  \[
  \upsilon^*_{N,M}(f)=\var(X_{f_1}) +\frac{y_{N}^2}{(y_{N}+y_{M})^2}\cdot \var(X_{f_2})
  -\frac{2y_{N}}{y_{N}+y_{M}}\cdot \cov(X_{f_1},X_{f_2})~,
  \]
  by plugging in the calculations above, we readily find the announced
  formula for
  $a_{N,M}^*(f)$ and $\upsilon^*_{N,M}(f)$.

\subsection{Proof of Theorem \ref{th:substitu1}}
\label{pf:substitu1}

The strategy of the proof and many basic
steps follow the proof of the CLT in \citet{BS04} so that we
emphasize on those calculations needed by the refinement proposed.
First of all, the truncation and the follow-up centering and
normalization steps are exactly the same, that is, under the
assumptions made on their moments, the variables $\{X_{ij}\}$'s
can be truncated at level $\eta_n\sqrt{n}$ without altering the
limiting spectral distribution, where $\eta_n\to 0$ slowly. Note
that the 4th moments of the truncated and re-normalized random
variables may not be the same but they will be of form
$\kappa+1+\beta_x+o(1)$, and for the complex case we have
$EX_{ij}^2=o(n^{-1})$. The support set of the LSD of $\bS_x$
is
$$[(1-\sqrt{y})^2I_{(0<y<1)}\lim\inf\limits_{n}\lambda_{\min}^{\bSi_x},~
(1+\sqrt{y})^2\lim\sup\limits_{n}\lambda_{\max}^{\bSi_x}].$$ Let
$x_r$ be a number greater than
$(1+\sqrt{y})^2\lim\sup\limits_{n}\lambda_{\max}^{\bSi_x}$. If
$y<1$, then let $x_l$ be a number between $0$ and
$(1-\sqrt{y})^2\lim\inf\limits_{n}\lambda_{\min}^{\bSi_x}$. If
$y\geq1$, let $x_l$ be a negative number. Let $\eta_l$ and
$\eta_r$ satisfy
$$x_l<\eta_l<(1-\sqrt{y})^2I_{(0<y<1)}\lim\inf\limits_{n}\lambda_{\min}^{\bSi_x}<
(1+\sqrt{y})^2\lim\sup\limits_{n}\lambda_{\max}^{\bSi_x}<\eta_r<x_r.$$
Define a contour
${\cal{C}}={\cal{C}}_l\cup{\cal{C}}_u\cup{\cal{C}}_b\cup{\cal{C}}_r$
where \bqn &&   {\cal{C}}_u=\{x+i\nu_0: x\in[x_l,x_r]\},  \quad
{\cal{C}}_l=\{x_l+i\nu: |\nu|\le \nu_0\},\\
&& {\cal{C}}_b=\{x-i\nu_0: x\in[x_l,x_r]\}, \quad
{\cal{C}}_r=\{x_r+i\nu: |\nu|\le \nu_0\}, \eqn and ${\cal
C}_n={\cal C}\bigcap\{z:\Im(z)>n^{-2}\}$. As $f_j$ is analytic, we
have by Cauchy integral theorem
\begin{eqnarray*}
  Y_p(f_j)&=&\sum\limits_{i=1}^pf_j(\lambda_i)-p\int
  f_j(x)dF^{y_{N},H_p}(x)\\
  &=&-\frac{1}{2\pi i}\oint\limits_{{\cal C}}
  f_j(z)\cdot\left(\frac{1}{p}\rtr(\bS_x-z{\bf
      I}_p)^{-1}-m_{N}^{(0)}(z)\right)dz,
\end{eqnarray*}
(see (1.14) of \citet{BS04} where $\underline
m_{N}^{(0)}(z)=-\frac{1-y_{N}}{z}+y_{N}m_{N}^{(0)}(z)$
with $y_{N}=p/N$ and
\begin{equation*}
  z=-\frac{1}{\underline{m}_{N}^{(0)}(z)}+\frac{p}{N}\int\frac{t}{1+t\underline{m}_{N}^{(0)}(z)}dH_p(t).
\end{equation*}
It remains to find the asymptotic distribution of
$$
\rtr(\bS_x-z{\bf I}_p)^{-1}-pm_{N}^{(0)}(z),
$$
in order to obtain the asymptotic distribution of $Y_p(f_j)$.
Define $\bgma_i=\frac{1}{\sqrt{n}}\bm{\Gamma} {X}_{i}$. Then,
we have
$$
{\bf S}_x=\frac{n}{N}\sum_{i=1}^n(\bgma_i-\bar
\bgma)(\bgma_i-\bar \bgma)^*
=\sum_{i=1}^n\bgma_i\bgma_i^*-\frac{1}{N}\sum_{i\ne
  j}\bgma_i\bgma_j^* ={\bf B}_x-\bm{\gD},
$$
where $\bar{\bgma}=\frac1n\sum\limits_{i=1}^n\bgma_i$,
$\bm{\Delta}=\frac1{N}\sum_{j\ne k}\bgma_j\bgma_k^*$ and
 ${\bf B}_x = \sum_{i=1}^n\bgma_i\bgma_i^*$.
Let  ${\bf A}(z)={\bf B}_x-z{\bf I}$.
We have,
\begin{eqnarray*}
  &&\left({\bf S}_x-z{\bf
      I}\right)^{-1}=\left(\bbA(z)-\bm{\gD}\right)^{-1}=\bbA^{-1}(z)
  +\left(\bbA(z)-\bm{\gD}\right)^{-1}\bm{\gD}\bbA^{-1}(z)\\
  &=&\bbA^{-1}(z)+\bbA^{-1}(z)\bm{\gD}\bbA^{-1}(z)+\bbA^{-1}(z)(\bm{\gD}\bbA^{-1}(z))^2
  +\left(\bbA(z)-\bm{\gD}\right)^{-1}\left(\bm{\gD}\bbA^{-1}(z)\right)^{3}.
\end{eqnarray*}
Therefore,
\begin{eqnarray}
  \nonumber    &&p\left(\frac{1}{p}{\rm tr}({{\bf S}_x}-z{\bf I})^{-1}-m_{N}^{(0)}(z)\right)\\
  \nonumber    &=&p\left(\frac{1}{p}{\rm tr}(\bbA(z)-\bm{\gD})^{-1}-m_{n}^{(0)}(z)+m_{n}^{(0)}(z)-m_{N}^{(0)}(z)\right)\\
  \nonumber    &=&p\left(\frac{1}{p}{\rm tr}{\bf
     A}^{-1}(z)-m_{n}^{(0)}(z)\right)+p(m_{n}^{(0)}(z)-m_{N}^{(0)}(z))
  +{\rm tr}{\bf A}^{-2}(z)\bm{\gD}\\
  &&+{\rm tr}{\bf A}^{-1}(z)(\bm{\gD}{\bf A}^{-1}(z))^2+{\rm tr}\left({\bf A}(z)-\bm{\gD}\right)^{-1}(\bm{\gD}{\bf A}^{-1}(z))^3
  \label{sn},
\end{eqnarray}
where $\underline m_n^{(0)}(z)=-\frac{1-y_n}{z}+y_nm_n^{(0)}(z)$,
$\underline m_n^{(0)}(z)$ and $\underline m_{N}^{(0)}(z)$
satisfy
\begin{equation}\label{eq1}
  z=-\frac{1}{\underline{m}_{n}^{(0)}(z)}+\frac{p}{n}\int\frac{t}{1+t\underline{m}_{n}^{(0)}(z)}dH_p(t),
\end{equation}

\begin{equation}\label{eq2}
  z=-\frac{1}{\underline{m}_{N}^{(0)}(z)}+\frac{p}{N}\int\frac{t}{1+t\underline{m}_{N}^{(0)}(z)}dH_p(t),
\end{equation}

\begin{equation}\label{eq12}
  z=-\frac{1}{\underline{m}_{y}(z)}+{y}\int\frac{t}{1+t\underline{m}_{y}(z)}dH(t).
\end{equation}

By (\ref{sn}) and Lemmas \ref{lem1} and \ref{thm2}, we have

\begin{eqnarray}
&&  \rtr({{\bf S}_x}-z{\bf I}_p)^{-1}-p\cdot
 m_{N}^{(0)}(z)=p\left(\frac{1}{p} {\rm tr}{\bf
      A}^{-1}(z)-m_{n}^{(0)}(z)\right)+o_p(1)\nonumber\\
  &&=p\left(\frac{1}{p} {\rm tr}({\bf B}_x-z{\bf
      I}_p)^{-1}-m_{n}^{(0)}(z)\right)+o_p(1).\label{end}
\end{eqnarray}

We also need to check the following two properties of tightness and
equicontinuity before concluding. \\[1mm]

\noindent {\bf (1) Tightness of $\rtr({{\bf S}_x}-z{\bf
    I}_p)^{-1}-pm_{N}^{(0)}(z)$.} Because $\rtr({{\bf S}_x}-z{\bf
  I}_p)^{-1}-pm_{N}^{(0)}(z)=\rtr({{\bf S}_x}-z{\bf
  I}_p)^{-1}-\rtr({\bf B}_x-z{\bf I}_p)^{-1}+\rtr({\bf B}_x-z{\bf
  I}_p)^{-1}-pm_{n}^{(0)}(z)+pm_{n}^{(0)}(z)-pm_{N}^{(0)}(z)$ and
the tightness of $\{\rtr({\bf B}_x-z{\bf
  I}_p)^{-1}-pm_{n}^{(0)}(z)\}$ proved in \citet{BS04},
then we only prove the tightness of
$\{\rtr({{\bf   S}_x}-z{\bf I}_p)^{-1}-\rtr({\bf B}_x-z{\bf I}_p)^{-1}\}$. Let
$\{\tilde{\lambda}_i\}$ and $\{\lambda_i\}$ be the eigenvalues of
$\bbB_x$ and $\bbS_x$ respectively and be arranged in descending
order. Let the even ${\cal B}_n$ is defined as $\eta_l<
\lambda_p<\frac{n}{N}\tilde{\lambda}_1<\eta_r$. Then it is
well-known from random matrix theory that
for any positive number $t$, it holds for large enough $n$ that
$$
{\rm P}({\cal B}_n^c)=o(n^{-t}),
$$
see e.g. \citet{BS98}.
Notice that
$\bS_x=\frac{n}{N}\bbB_x-\frac{n}{N}\bm{\Gamma}\bar{{
    X}}\bar{{X}}^*\bm{\Gamma}_x^*$ where
$\bar X=\sum_{j=1}^n X_j$.  Similar to arguments in
\citet{BS04}, we only need to prove that there is an absolute constant
$M$ such that for any $z_1,z_2\in {\cal C}_n$, \bqa
&&\frac{E|\rtr(\bbB_x-z_1\bbI_p)^{-1}-\rtr(\bbB_x-z_2\bbI_p)^{-1}-\rtr
  (\bbS_x-z_1\bbI_p)^{-1}+\rtr(\bbS_x-z_2\bbI_p)^{-1}|^2}{|z_1-z_2|^2}\nonumber\\
&=&\rE\Big|\rtr(\bbB_x-z_1\bbI_p)^{-1}(\bbB_x-z_2\bbI_p)^{-1}-\rtr
(\bbS_x-z_1\bbI_p)^{-1}
(\bbS_x-z_2\bbI_p)^{-1}\Big|^2\nonumber\\
&=&\rE\left|\sum_{i=1}^n\frac{(\tilde\lambda_i-\lambda_i)(\lambda_i+\tilde\lambda_i-z_1-z_2)}{(\lambda_i-z_1)
    (\lambda_i-z_2)(\tilde\lambda_i-z_1)(\tilde\lambda_i-z_2)}\right|^2\nonumber\\
&\le&
K\rE\left\{ \left|\sum_{i=1}|\lambda_i-\tilde\lambda_i|\right|^2
  I_{{\cal  B}_n}\right\}+o(1)\le M,\label{eq36}
\eqa
where the last step of
(\ref{eq36}) follows from the fact that
\begin{eqnarray*}
  \sum\limits_{i=1}|\lambda_i-\tilde\lambda_i|I_{{\cal  B}_n}
  &=&\sum\limits_{i=1}|\lambda_i-\frac{n}{N}\tilde\lambda_i|
  +\frac{1}{N}\sum\limits_{i=1}^n\tilde\lambda_i\\
  &=&\sum_{i=1}\frac{n}{N}\tilde\lambda_i-\lambda_i+\frac{1}{N}\sum\limits_{i=1}^n\tilde\lambda_i\\
  &\le&\frac{n}{N}\tilde{\lambda}_1-\lambda_p+\frac{1}{N}\sum\limits_{i=1}^n\tilde\lambda_i\\
  &\le&2\eta_r-\eta_l,
\end{eqnarray*}
by the interlacing theorem.

\noindent {\bf (2) The equi-continuity of $\rE\rtr({{\bf S}_x}-z{\bf
    I})^{-1}-pm^{(0)}_{N}(z)$} can be proved in a similar way to
that for the tightness of $\rtr({{\bf S}_x}-z{\bf
  I})^{-1}-\rE\rtr({{\bf B}_x}-z{\bf I})^{-1}$, see \cite{BS04} and \cite{Zheng2012}.

Finally, the proof is completed.

\subsection{Proof of Theorem~\protect\ref{th:newCLT}}\label{pf:newCLT}

Now we assume that the matrix $\bm{\Gamma}$ is real, $|EX_{11}^2|<\infty$. Then we consider
(2.7) of Bai and Silverstein (2004). According to Bai and Silverstein (2004), it is easy to obtain
\bqn
\frac{1}{n}\rE_j[z_1\rtr\bm{\Sigma}_x\bbA_j^{-1}(z_1)-z_2\rtr\Sigma_x(\breve\bbA_j')^{-1}(z_2)]\to z_1(b^{-1}(z_1)-1)-z_2(b^{-1}(z_2)-1),a.s.
\eqn
where $\rE_j$ is the conditional expectation on $\bbr_1,\ldots,\bbr_{j-1}$, $\breve\bbr_{j+1},\ldots,\breve\bbr_{n}$
are an independent copy of $\bbr_{j+1},\ldots,\bbr_n$,
$\bbA_j=\sum\limits_{i=1}^{n}\bbr_i\bbr_i^{*}-z{\bf I}_p-\bbr_j\bbr_j^{*}$, $\bbA_{kj}=\sum\limits_{i=1}^{n}\bbr_i\bbr_i^{*}-z{\bf I}_p-\bbr_j\bbr_j^{*}-\bbr_k\bbr_k^{*}$, $\bbr_j=\bm{\Gamma}{\bf X}_{\cdot j}$,
$\bm{\beta}_{k}=\frac{1}{1+\bbr_k^{*}\bbA_k^{-1}\bbr_k}$, $\bm{\beta}_{k(j)}=\frac{1}{1+\bbr_k^{*}\bbA_{kj}^{-1}\bbr_k}$
 and $b(z)\rightarrow-z\underline{m}_{y}(z)$.
Moreover,
\bqn
&&\frac1n[z_1\rE_j\rtr\bm{\Sigma}_x\bbA_j^{-1}(z_1)-z_2\rtr\Sigma_x(\breve\bbA_j')^{-1}(z_2)]\\
&=&
\frac1n\rE_j\rtr\Sigma_x\bbA_j^{-1}(z_1)\Big[\sum_{i=1}^{j-1}(z_1\bar\bbr_i\bbr_i'-z_2\bbr_i\bbr_i^*)+
\sum_{i=j+1}^n(
z_1\bar{\breve\bbr}_i\breve\bbr_i'-z_2\bbr_i\bbr_i^*)\Big](\breve\bbA_j')^{-1}(z_2)\\
&=&
\frac1n\sum_{i=1}^{j-1}\rE_j\bigg\{z_1\breve\beta_{ji}(z_2)\bbr_i'(\breve\bbA_{ji}')^{-1}(z_2)\Sigma_x\left[\bbA_{ji}^{-1}(z_1)
-\bbA_{ji}^{-1}(z_1)\bbr_i\bbr_i^*\bbA_{ji}^{-1}(z_1)\beta_{ji}(z_1)\right]\bar\bbr_i\\
&&-z_2\bbr_i^*\left[(\breve\bbA_{ji}')^{-1}(z_2)-\breve\beta_{ji}(z_2)(\breve\bbA'_{ji})^{-1}(z_2)\bar\bbr_i\bbr_i'
(\breve\bbA'_{ji})^{-1}(z_2)\right]
\Sigma_x\bbA_{ji}^{-1}(z_1)
\beta_{ji}(z_1)\bbr_i\bigg\}\\
&&+\frac1n\sum_{i=j+1}^n\rE_j\Big[
z_1\breve\beta_{ji}(z_2)\breve\bbr_i'(\breve\bbA_{ji}')^{-1}(z_2)\Sigma_x\bbA_j^{-1}(z_1)\bar{\breve\bbr}_i-
z_2\bbr_i^*(\breve\bbA_{j}')^{-1}(z_2)\Sigma_x\bbA_{ji}^{-1}(z_1)\bbr_i\beta_{ji}(z_1)\Big]
\eqn
\bqn
&=&
\frac{1}{n}\sum_{i=1}^{j-1}\rE_j\bigg\{z_1\breve\beta_{ji}(z_2)\left(\frac{1}{n}\rtr(\breve\bbA_{ji}')^{-1}(z_2)\Sigma_x\bbA_{ji}^{-1}(z_1)\Sigma_x
-\frac{\alpha_x}{n^2}\rtr(\breve\bbA_{ji}')^{-1}(z_2)\Sigma_x\bbA_{ji}^{-1}(z_1)\Sigma_x\cdot\rtr\bbA_{ji}^{-1}(z_1)\Sigma_x\beta_{ij}(z_1)
\right)\\
&&-z_2\beta_{ij}(z_1)\left(\frac{1}{n}\rtr(\breve\bbA_{ji}')^{-1}(z_2)\Sigma_x\bbA_{ji}^{-1}(z_1)\Sigma_x-
\frac{\alpha_x}{n^2}\rtr(\breve\bbA'_{ji})^{-1}(z_2)\Sigma_x\cdot\rtr
(\breve\bbA'_{ji})^{-1}(z_2)
\Sigma_x\bbA_{ji}^{-1}(z_1)\Sigma_x\breve\beta_{ji}(z_2)\right)\\
&&+\frac1n\sum_{i=j+1}^n\rE_j\Big[
z_1\breve\beta_{ji}(z_2)\breve\bbr_i'(\breve\bbA_{ji}')^{-1}(z_2)\Sigma_x\bbA_j^{-1}(z_1)\bar{\breve\bbr}_i-
z_2\bbr_i^*(\breve\bbA_{j}')^{-1}(z_2)\Sigma_x\bbA_{ji}^{-1}(z_1)\bbr_i\beta_{ji}(z_1)\Big]+o_{a.s.}(1)\\
&=&\left\{-\frac{j-1}n\alpha_x[(z_1b(z_2)-z_2b(z_1))-b(z_1)b(z_2)(z_1-z_2)]+(z_1b(z_2)-z_2b(z_1))\right\}\\
&&\rE_j\frac1n\rtr\Sigma_x\bbA_{j}^{-1}(z_1)\Sigma_x(\breve\bbA_{j}')^{-1}(z_2)+o_{a.s.}(1)
\eqn
where $\alpha_x=|EX_{11}^2|^2$.
Comparing the two estimates, we obtain
$$
\rE_j\frac1n\rtr\Sigma_x\breve\bbA_{j}^{-1}(z_2)\Sigma_x\bbA_{j}^{-1}(z_1)=\frac{z_1(b^{-1}(z_1)-1)-(z_2b^{-1}(z_2)-1)+o_{a.s.}(1)}
{-\frac{j-1}n\alpha_x[(z_1b(z_2)-z_2b(z_1))-b(z_1)b(z_2)(z_1-z_2)]+(z_1b(z_2)-z_2b(z_1))}
$$
Consequently, we obtain
\begin{eqnarray*}
&&\frac1{n^2}\sum_{j=1}^n\alpha_xb_n(z_1)b_n(z_2)\rtr\rE_{j}\Sigma_x\bbA^{-1}_j(z_1)
\rE_j(\Sigma_x(\bbA'_j)^{-1}(z_2))\\
&\to&a(z_1,z_2)\int_0^1\frac1{1-ta(z_1,z_2)}dt=\log(1-a(z_1,z_2)),
\end{eqnarray*}
where
\begin{eqnarray}
  \nonumber
  a(z_1,z_2)&=&\frac{\alpha_xb(z_1)b(z_2)(z_1(b^{-1}(z_1)-1)-z_2(b^{-1}(z_2)-1))}{z_1b(z_2)-z_2b(z_1)}\\
  \nonumber &=&\alpha_x(1+\frac{b(z_1)b(z_2)(z_2-z_1)}{z_1b(z_2)-z_2b(z_1)})\\
  &=&\alpha_x\left(1+\frac{\underline{m}_{y}(z_1)\underline{m}_{y}(z_2)(z_1-z_2)}{\underline{m}_{y}(z_2)-\underline{m}_{y}(z_1)}\right).
  \label{eq:az1z2}
\end{eqnarray}
Moreover we have
$$
\frac{d^2}{d z_2d z_1}\int_0^{a(z_1,z_2)} \frac1{1-z}dz=\frac{d^2}{d z_1d z_2}\log(1-a(z_1,z_2)).
$$
So the covariance function $Cov(X_{f_1}, X_{f_2})$ will have an additional term as follows
\begin{equation}\label{acov}
\frac{-1}{4\pi^2}\oint\oint f_1(z_1)f_2(z_2)\left[\frac{d^2}{d z_1d z_2}\log(1-a(z_1,z_2))\right]dz_1dz_2.
\end{equation}
By (4.10) of Bai and Silverstein (2004), we have
\bqn
&&yn^{-1}\sum\limits_{j=1}^n\rE\beta_jd_j\\
&=&\frac{z^2\underline{m}_{y}^2(z)}{n^2}\sum\limits_{j=1}^{n}\rE\rtr\bbA_j^{-1}(z)(\underline{m}_{y}(z)\Sigma_x+{\bf I}_p)^{-1}\Sigma_x\bbA_j^{-1}(z)\Sigma_x+o(1)\\
&=&\frac{\underline{m}_{y}^2(z)}{n}\rE\rtr(\underline{m}_{y}(z)\Sigma_x+{\bf I}_p)^{-3}\Sigma_x^2\\
&&+\frac{z^2\underline{m}_{y}^4(z)}{n^2}\sum\limits_{i\not=j}
\rE\rtr(\underline{m}_{y}(z)\Sigma_x+{\bf I}_p)^{-2}\Sigma_x(\bbr_i\bbr_i^{*}-\frac{1}{n}\Sigma_x)\bbA_{ij}^{-1}(z)(\underline{m}_{y}(z)\Sigma_x+{\bf I}_p)^{-1}\bbA_{ij}^{-1}(z)(\bar\bbr_i\bbr_i'-\frac{1}{n}\Sigma_x)+o(1)\\
&=&\frac{\underline{m}_{y}^2(z)}{n}\rE\rtr(\underline{m}_{y}(z)\Sigma_x+{\bf I}_p)^{-3}\Sigma_x^2\\
&&+\frac{\alpha_xz^2\underline{m}_{y}^4(z)}{n^3}\sum\limits_{j=1}^{n}\rE\rtr(\underline{m}_{y}(z)\Sigma_x+{\bf I}_p)^{-2}\Sigma_x^2\cdot\rtr\bbA_j^{-1}(z)(\underline{m}_{y}(z)\Sigma_x+{\bf I}_p)^{-1}\bbA_{j}^{-1}(z)\Sigma_x+o(1).
\eqn
Then we have
\begin{eqnarray*}
y_1n^{-1}\sum\limits_{j=1}^nE\beta_jd_j
&=&\frac{\alpha_x\frac{\underline{m}_{y}^2(z)}{n}\rE\rtr(\underline{m}_{y}(z)\Sigma_x+{\bf I}_p)^{-3}\Sigma_x^2}{1-\frac{\alpha_x\underline{m}_{y}^2(z)}{n}\rtr(\underline{m}_{y}(z)\Sigma_x+{\bf I}_p)^{-2}\Sigma_x^2}+o(1)\\
&=&\frac{\alpha_xy_1\int\frac{\underline{m}_{y}^2(z)t^2dH(t)}{(1+t\underline{m}_{y}(z))^3}}{1-\alpha_xy_1\int\frac{\underline{m}_{y}^2(z)t^2dH(t)}{(1+t\underline{m}_{y}(z))^2}}+o(1).
\end{eqnarray*}
Then the mean function $\rE X_f$ of Bai and Silverstein (2004) will be
\begin{equation}\label{amean}
\frac{1}{2\pi i}\oint\frac{\alpha_xy\int\frac{\underline{m}_{y}^3(z)t^2dH(t)}{(1+t\underline{m}_{y}(z))^3}}{\left(1-y_1\int\frac{\underline{m}_{y}^2(z)t^2dH(t)}{(1+t\underline{m}_{y}(z))^2}\right)\left(1-\alpha_xy_1\int\frac{\underline{m}_{y}^2(z)t^2dH(t)}{(1+t\underline{m}_{y}(z))^2}\right)}dz.
\end{equation}
By (\ref{end}), Zheng (2012)'s (40)-(41) (or Pan (2012)) and Bai and Silverstein
(2004)'s Theorem 1.1, we have we can show that
\begin{eqnarray*}
  X_p(f_j)&=&\sum\limits_{i=1}^pf_j(\lambda_i)-p\int
  f_j(x)dF^{y_{n},H_p}(x)\\
  &=& \oint f_j(z)[\rtr({{\bf S}_x^0}-z{\bf I}_p)^{-1}-p\cdot
  m_{n}^{(0)}(z)]dz,
\end{eqnarray*}
converges to a Gaussian vector with mean $EX_{f_j}$ and covariance
function $\cov(X_{f_j}, X_{f_k})$ as follows
\begin{eqnarray}
  EX_f&=&-\frac{1}{2\pi i}\oint
  f(z)\frac{y_1\alpha_x\oint\frac{\underline{m}_{y}^3(z)t^2}{(1+t\underline{m}_{y}(z))^{3}}dH(t)}
  {\left(1-y_1\oint\frac{\underline{m}_{y}^2(z)t^2}{(1+t\underline{m}_{y}(z))^{2}}dH(t)\right)\left(1-y_1\alpha_x\oint\frac{\underline{m}_{y}^2(z)t^2}{(1+t\underline{m}_{y}(z))^{2}}dH(t)\right)}dz\nonumber \\
  &&-\frac{y_1(E|X_{11}^4|-\alpha_x-2)}{2\pi i}\cdot\int f(z)\frac{\underline{m}_{y}^3(z)M(z)}
  {1-y_1\int\frac{\underline{m}_{y}^2(z)t^2dH(t)}
    {(1+t\underline{m}_{y}(z))^2}}dz\\ \nonumber
\end{eqnarray}
and
\begin{eqnarray}
  \mbox{Cov}(X_f,
  X_g)&=&-\frac{1}{4\pi^2}\oint\oint\frac{f(z_1)g(z_2)}{(\underline{m}_{y}(z_1)-\underline{m}_{y}(z_2))^2}
  \frac{d}{dz_1}\underline{m}_{y}(z_1)\frac{d}{dz_2}\underline{m}_{y}(z_2) dz_1dz_2\nonumber\\
  &&-\frac{y(E|X_{11}^4|-\alpha_x-2)}
  {4\pi^2}\int\int
  f_1(z_1)f_2(z_2)\frac{d^2}{dz_1dz_2}\underline{m}_{y}(z_1)\underline{m}_{y}(z_2)
  C(z_1,z_2)dz_1dz_2\nonumber\\
  &&-\frac{1}{4\pi^2}\oint\oint f_1(z_1)f_2(z_2)\left[\frac{d^2}{d z_1d z_2}\log(1-a(z_1,z_2))\right]dz_1dz_2\nonumber
\end{eqnarray}
where $a(z_1,z_2)$ is given in \eqref{eq:az1z2},
$M(z)$ be the limit of
\[ \frac1p\sum\limits_{i=1}^p{\bf
  e}_i^*\bm{\Gamma}_x^*(\underline{m}_{y}(z)\bm{\Sigma}_x+{\bf
  I}_p)^{-1}\bm{\Gamma}_x{\bf e}_i\cdot{\bf
  e}_i^*\bm{\Gamma}_x^*(\underline{m}_{y}(z)\bm{\Sigma}_x+{\bf
  I}_p)^{-2}\bm{\Gamma}_x{\bf e}_i,
\]
and $C(z_1,z_2)$  the limit
of
\[ \frac1p\sum\limits_{i=1}^p{\bf
  e}_i^*\bm{\Gamma}_x^*(\underline{m}_{y}(z_1)\bm{\Sigma}_x+{\bf
  I}_p)^{-1}\bm{\Gamma}_x{\bf e}_i\cdot{\bf
  e}_i^*\bm{\Gamma}_x^*(\underline{m}_{y}(z_2)\bm{\Sigma}_x+{\bf
  I}_p)^{-1}\bm{\Gamma}_x{\bf e}_i,
\]
with the $i$-th unit vector $\mathbf{e}_i$
(null coordinates except  the $i$th equal to 1).
Suppose that
$\bm{\Gamma}^{*}\bm{\Gamma}$ is diagonal with eigenvalues
$\{\lambda_{j0}\}$. It follows that
$$
{\bf
  e}_i^*\bm{\Gamma}_x^*(\underline{m}_{y}(z)\bm{\Sigma}_x+{\bf
  I}_p)^{-1}\bm{\Gamma}_x{\bf e}_i=\frac{\lambda_{i0}}{\underline
  m_{y}(z)\lambda_{i0}+1},
$$
and
$$
{\bf
  e}_i^*\bm{\Gamma}_x^*(\underline{m}_{y}(z)\bm{\Sigma}_x+{\bf
  I}_p)^{-2}\bm{\Gamma}_x{\bf e}_i=\frac{\lambda_{i0}}{(\underline
  m_{y}(z)\lambda_{i0}+1)^2}.
$$
Thus, it follows that
\begin{eqnarray*}
  &&\frac1p\sum\limits_{i=1}^p{\bf
    e}_i^*\bm{\Gamma}_x^*(\underline{m}_{y}(z)\bSi_{x}+{\bf
    I}_p)^{-1}\bm{\Gamma}_x{\bf e}_i\cdot{\bf
    e}_i^*\bm{\Gamma}_x^*(\underline{m}_{y}(z)\bSi_x +{\bf
   I}_p)^{-2}\bm{\Gamma}_x{\bf
    e}_i\\
  &=&\frac1p\sum\limits_{i=1}^p\frac{(\lambda_{i0})^2}{(\underline
    m_{y}(z)\lambda_{i0}+1)^3}\stackrel{}{\rightarrow}
  \int\frac{t^2}{(\underline m_{y}(z)t+1)^3}dH(t),
\end{eqnarray*}
and
\begin{eqnarray}
  &&\frac1p\sum\limits_{i=1}^p{\bf
    e}_i^*\bm{\Gamma}_x^*(\underline{m}_{y}(z_1)\bSi_x+{\bf
    I}_p)^{-1}\bm{\Gamma}_x{\bf e}_i\cdot{\bf
    e}_i^*\bm{\Gamma}_x^*(\underline{m}_{y}(z_2)\bSi_x
  +{\bf I}_p)^{-1}\bm{\Gamma}_x{\bf e}_i\nonumber\\
  &&\stackrel{p}{\rightarrow}\int\frac{t^2}{(\underline
    m_{y}(z_1)t+1)(\underline m_{y}(z_2)t+1)}dH(t)\nonumber
\end{eqnarray}
because of $\bm{\Gamma}^*\bm{\Gamma}$ is diagonal. So we obtain
\begin{eqnarray}
  \rE X_f&=&-\frac{1}{2\pi i}\oint
  f(z)\frac{\alpha_xy_1\int\frac{\underline{m}_{y}^3(z)t^2}{(1+t\underline{m}_{y}(z))^{3}}dH(t)}
  {\left(1-y_1\int\frac{\underline{m}_{y}^2(z)t^2}{(1+t\underline{m}_{y}(z))^{2}}dH(t)\right)\left(1-\alpha_xy_1\int\frac{\underline{m}_{y}^2(z)t^2}{(1+t\underline{m}_{y}(z))^{2}}dH(t)\right)}dz\nonumber\\
  &&-\frac{(E|X_{11}^4|-\alpha_x-2)}{2\pi i}\cdot\oint\left[f(z)\cdot\frac{y_1\underline{m}_{y}^3(z)\int\frac{t^2}
      {(\underline m_{y}(z)t+1)^3}dH(t)}{1-y_1\int\frac{\underline{m}_{y}^2(z)t^2dH(t)}
      {(1+t\underline{m}_{y}(z))^2}}\right]dz,\nonumber
\end{eqnarray}
and
\begin{eqnarray}
  \mbox{Cov}(X_f,
  X_g)&=&-\frac{1}{4\pi^2}\oint\oint\frac{f(z_1)g(z_2)}{(\underline{m}_{y}(z_1)-\underline{m}_{y}(z_2))^2}
  \frac{d}{dz_1}\underline{m}_{y}(z_1)\frac{d}{dz_2}\underline{m}_{y}(z_2) dz_1dz_2\label{covA3}\\
  &&-\frac{y_1(E|X_{11}^4|-\alpha_x-2)}{4\pi^2}\oint\oint\left[\int\frac{f_1(z_1)}{(\underline m_{y}(z_1)t+1)^2}
    \frac{f_2(z_2)}{(\underline m_{y}(z_1)t+1)^2}dH(t)\right]dz_1dz_2\nonumber\\
  &&-\frac{1}{4\pi^2}\oint\oint f_1(z_1)f_2(z_2)\left[\frac{d^2}{d z_1d z_2}\log(1-a(z_1,z_2))\right]dz_1dz_2
\end{eqnarray}
The proof of Theorem~\ref{th:newCLT} is complete.

\subsection{Proof of Theorem~\protect\ref{th:substitu2}}
\label{pf:substitu2}

Recall that $N=n-1$ and $M=m-1$ are the adjusted sample sizes.
The proof has two steps following the decomposition
$$
\rtr({\bf F}-z{\bf I}_p)^{-1}-pm_{(y_{N},y_{M})}(z)=
\left[\rtr({\bf S}_x{\bf S}_y^{-1}-z{\bf
I}_p)^{-1}-pm^{(y_{N},F^{{\bf S}^{-1}_y})}(z)\right]
$$
$$
+p\left[m^{(y_{N},F^{{\bf
S}^{-1}_y})}(z)-m_{(y_{N},y_{M})}(z)\right],
$$
where
\begin{itemize}
\item
  $F^{{\bf S}_y^{-1}}(t)$ and $F^{{\bf S}_y}(t)$ are the ESDs
  of ${\bf S}_y^{-1}$ and ${\bf S}_y$;
\item
  $m_{(y_1, y_2)}(z)$ is the
  Stieltjes transform of the LSD $G_{(y_1,y_2)}$ of $\bF$,
  and to simplify the notations we simply write
  $m(z) = m_{(y_1, y_2)}(z)$ if no confusion is possible, and
  $\underline{m}(z)=-\frac{1-y_1}{z}+y_1m(z)$;
\item
  $m_{(y_{N},y_{M})}(z)$ is obtained by replacing $(y_1, y_2)$
  by $(y_{N},y_{M})$ in $m(z)=m_{(y_1, y_2)}(z)$;  and
\item
  $$
  \underline{m}^{\{y_{N},F^{{\bf
        S}_y^{-1}}\}}=-\frac{1-y_{N}}{z}+ y_{N}m^{\{y_{N},F^{{\bf
        S}_y^{-1}}\}}(z)
  $$
  so that we have
  \bqa z&=&-\frac{1}{\underline{m}^{\{y_{N},F^{{\bf S}_y^{-1}}\}}}
  +y_{N}\int\frac{t}{1+t\underline{m}^{\{y_{N},F^{{\bf S}_y^{-1}}\}}}dF^{{\bf S}_y^{-1}}(t)\nonumber\\
  &=&-\frac{1}{\underline{m}^{\{y_{N},F^{{\bf S}_y^{-1}}\}}}
  +y_{N}\int\frac{1}{t+\underline{m}^{\{y_{N},F^{{\bf
          S}_y^{-1}}\}}}dF^{{\bf S}_y}(t).\label{eq25} \eqa
\end{itemize}
\medskip\noindent{\bf Step 1.}\quad  Given
${\bf S}_y$, in the proof of Theorem \ref{th:substitu1}, we have
proved that the process $\{\rtr({\bf S}_x{\bf S}_y^{-1}-z{\bf
I}_p)^{-1}-pm^{\{y_{N},F^{{\bf S}_y^{-1}}\}}(z)\}$ weakly tends
to a Gaussian process $M_1(z)$ on the contour ${\mathcal C}$ with
mean function
\begin{eqnarray*}
\rE\left(M_1(z)|\mathscr {\bf
S}_2\right)&=&\displaystyle{(\kappa-1)\cdot\frac{
y_1\int\underline{m}(z)^3
x[x+\underline{m}(z)]^{-3}dF_{y_{2}}(x)}{\left[1-y_1\int
\underline{m}^2(z)(x+\underline{m}(z))^{-2}dF_{y_{2}}(x)\right]^2}}\\
  &&+\displaystyle{\beta_x
\cdot\frac{y_1\cdot\underline{m}^3(z)\cdot\int\frac{dF_{y_2}(x)}{x+\underline{m}(z)}
\int\frac{x\cdot dF_{y_2}(x)}{(x+\underline{m}(z))^2}}
{1-y_1\int\underline{m}^2(z)(x+\underline{m}(z))^{-2}dF_{y_2}(x)}}
\end{eqnarray*}
for $z\in \mathcal{C}$ and covariance function \begin{eqnarray*}
\cov(M_1(z_1), M_1(z_2)|\mathscr {\bf S}_2)
&=&\displaystyle{\kappa\cdot\left(\frac{\underline{m}'(z_1)\cdot\underline{m}'(z_2)}{(\underline{m}(z_1)-
\underline{m}(z_2))^2}-\frac{1}{(z_1-z_2)^2}\right)}\\
&&+\displaystyle{\beta_x\cdot y_1\cdot
\int\frac{\underline{m}'(z_1)\cdot x\cdot
dF_{y_2}(x)}{(x+\underline{m}(z_1))^2}\int\frac{\underline{m}'(z_2)\cdot
x\cdot dF_{y_2}(x)}{(x+\underline{m}(z_2))^2}}
\end{eqnarray*}
for $z_1,z_2\in \mathcal{C}$ where $F_{y_2}$ is the
Mar$\check{c}$enko-Pastur law with the ratio $y_2$.

\medskip\noindent{\bf Step 2.}\quad  By (\ref{eq25}) and the truth of \bqa
z&=&-\frac{1}{\underline{m}_{\{y_{N},y_{M}\}}}
+y_{N}\int\frac{1}{t+\underline{m}_{\{y_{N},y_{M}\}}}dF_{y_{M}}(t),\label{eq26}
\eqa where $F_{y_{M}}$ is the Mar$\check{c}$enko-Pastur law with
the ratio $y_{M}=p/(M)$. Subtracting both sides of
(\ref{eq25}) from those of (\ref{eq26}) and by Theorem
\ref{th:substitu1}, we obtain \bqa\label{F}
&&{p\cdot\left[m^{\{y_{N},F^{{\bf
S}^{-1}_y}\}}(z)-m_{\{y_{N},y_{M}\}}(z)\right]}
\nonumber \\
&=&{-y_{N}\underline{m}_{\{y_{n-},y_{M}\}}\underline{m}^{\{y_{{N}},
F^{{\bf S}^{-1}_y}\}}\frac{\rtr\left({\bf
S}_{y}+\underline{m}_{\{y_{N},y_{M}\}}{\bf I}_p\right)^{-1}
-pm_{y_{M}}(-\underline{m}_{\{y_{N},y_{M}\}})}
{1-y_{N}\cdot\int\frac{\underline{m}_{\{y_{N},y_{M}\}}
\cdot\underline{m}^{\{y_{N},F^{{\bf S}^{-1}_y}\}}dF_{M}(t)}
{\left(t+\underline{m}_{\{y_{N},y_{M}\}}\right)\cdot
\left(t+\underline{m}^{\{y_{N},F^{{\bf S}^{-1}_x}\}}\right)}}}\nonumber\\
\eqa which converges weakly to a Gaussian process $M_2(\cdot)$ on
$z\in \mathcal{C}$ with mean function and covariance function
\begin{eqnarray*}
\rE(M_2(z))&=&-(\kappa-1)\cdot\frac{y_2\underline{m}'(z)\cdot
[\underline{m}_{y_2}(-\underline{m}(z))]^3\cdot[1+\underline{m}_{y_2}(-\underline{m}(z))]^{-3}}
{\left[1-y_2\cdot\left(\frac{\underline{m}_{y_2}(-\underline{m}(z))}{1+
\underline{m}_{y_2}(-\underline{m}(z))}\right)^2\right]^2}\\
&&\displaystyle{-\beta_y\cdot\underline{m}'(z)\frac{ y_2\cdot
m_0^3(z)\cdot(1+m_0(z))^{-3}} {1-y_2\cdot
m_0^2(z)\cdot(1+m_0(z))^{-2}}}
\end{eqnarray*}
and
\begin{eqnarray*}
&&\cov(M_2(z_1),
M_2(z_2))\\
&=&\kappa\underline{m}'(z_1)\underline{m}'(z_2)
\left(\frac{\underline{m}_{y_2}'(-\underline{m}(z_1))\cdot\underline{m}_{y_2}'
(-\underline{m}(z_2))}{[\underline{m}_{y_2}(-\underline{m}(z_1))-\underline{m}_{y_2}(-\underline{m}(z_2))]^2}-
\frac{1}{(\underline{m}(z_1)-\underline{m}(z_2))^2}\right)\\
&&\displaystyle{+\beta_y\cdot y_2\cdot
\frac{\underline{m}'(z_1)\underline{m}_{y_2}'(-\underline{m}(z_1))}
{(1+\underline{m}_{y_2}(-\underline{m}(z_1)))^2}
\cdot\frac{\underline{m}'(z_2)
\underline{m}_{y_2}'(-\underline{m}(z_2))}
{(1+\underline{m}_{y_2}(-\underline{m}(z_2)))^2}}
\end{eqnarray*}
for $z_1,z_2\in \mathcal{C}$,
\[z=-\frac{1}{\underline{m}_{y_{M}}}+\frac{y_{M}}{1+\underline{m}_{y_{M}}}, ~~
\underline{m}_{y_{M}}(z)=-\frac{1-y_{M}}{z}+y
m_{y_{M}}(z), \quad \text{with}~~
m_0(z)=\underline{m}_{y_2}(-\underline{m}(z_2)).
\]
Thus $\rtr({\bf F}-z{\bf I}_p)^{-1}-pm_{(y_{N},y_{M})}(z)$
converges to a Gaussian process $\{M_1(z)+M_2(z)\}$ with mean and
covariance functions as follows
\begin{eqnarray*}
\rE(M_1(z)+M_2(z))&=&
\displaystyle{(\kappa-1)\cdot\frac{y_1\int\underline{m}^3(z)
x[x+\underline{m}(z)]^{-3}dF_{y_2}(x)}{\left[1-y_1\int
\underline{m}^2(z)(x+\underline{m}(z))^{-2}dF_{y_2}(x)\right]^2}}\\
\\
&&\displaystyle{+\beta_x\cdot\frac{y_1\cdot\underline{m}^3(z)\cdot\int\frac{dF_{y_2}(x)}{x+\underline{m}(z)}
\int\frac{x\cdot dF_{y_2}(x)}{(x+\underline{m}(z))^2}}
{1-y_1\int\underline{m}^2(z)(x+\underline{m}(z))^{-2}dF_{y_2}(x)}}\\
\\
&&\displaystyle{-(\kappa-1)\cdot\underline{m}'(z)\frac{y_2\cdot
[m_0(z)]^3\cdot[1+m_0(z)]^{-3}}
{\left[1-y_2\cdot\left(\frac{m_0(z)}{1+
m_0(z)}\right)^2\right]^2}}\\
\\
&&\displaystyle{-\beta_y\cdot\underline{m}'(z)\frac{ y_2\cdot
m_0^3(z)\cdot(1+m_0(z))^{-3}} {1-y_2\cdot
m_0^2(z)\cdot(1+m_0(z))^{-2}}}
\end{eqnarray*}
and
\begin{eqnarray*}
&&\cov(M_1(z_1)+M_2(z_1),
M_1(z_2)+M_2(z_2))\\
\\
&=&\displaystyle{\beta_x\cdot y_1\cdot
\int\frac{\underline{m}'(z_1)\cdot x\cdot
dF_{y_2}(x)}{(x+\underline{m}(z_1))^2}\int\frac{\underline{m}'(z_2)\cdot
x\cdot
dF_{y_2}(x)}{(x+\underline{m}(z_2))^2}-\frac{\kappa}{(z_1-z_2)^2}}\\
\\
&&\displaystyle{+\kappa\cdot \frac{ m_0'(z_1)\cdot m_0'
(z_2)}{[m_0(z_1)-m_0(z_2)]^2}}\displaystyle{+\beta_y\cdot y_2\cdot
\frac{m_0'(z_1)} {(1+m_0(z_1))^2} \cdot\frac{ m_0'(z_2)}
{(1+m_0(z_2))^2}}.
\end{eqnarray*}
Then by Corollary 3.2 of Zheng (2012), we obtain that the random
vector $(W_p(f_1),\ldots,W_p(f_k))$ where
$$
W_p(f_j)=\sum\limits_{i=1}^pf_j(\lambda_i)-p\int
f_j(x)dF^{\{y_{N},y_{M}\}}(x)
$$
with eigenvalues $\lambda_i$ of  ${\bf F}$ converges weakly to
a Gaussian vector $(Z_{f_j})$ with mean
and covariance functions
\begin{eqnarray}
  &&\rE Z_{f_j}\nonumber\\
  &=&
  \lim_{r\downarrow 1}\frac{\kappa-1}{4\pi
    i}\ointctrclockwise\limits_{|\xi|=1}
  f_j\left(\frac{1+h^2+2h\Re(\xi)}{(1-y_2)^2}\right)
  \left[\frac{1}{\xi-r^{-1}}+\frac{1}{\xi+r^{-1}}-\frac{2}{\xi+\frac{y_2}{h}}\right]~~d\xi
\label{m1}\\
    && +\frac{\beta_xy_1(1-y_2)^2}{2\pi i\cdot
    h^2}\ointctrclockwise\limits_{|\xi|=1}
  f_j\left(\frac{1+h^2+2h\Re(\xi)}{(1-y_2)^2}\right)
  \frac{1}{(\xi+\frac{y_2}{h})^3}~~d\xi~,
  \nonumber\\
   && +\frac{\beta_y(1-y_2)}{4\pi
    i}\ointctrclockwise\limits_{|\xi|=1}
  f_j\left(\frac{1+h^2+2h\Re(\xi)}{(1-y_2)^2}\right)
  \frac{\xi^2-\frac{y_2}{h^2}}{(\xi+\frac{y_2}{h})^2}
  \left[\frac{1}{\xi-\frac{\sqrt{y_2}}{h}}
   +\frac{1}{\xi+\frac{\sqrt{y_2}}{h}}-\frac{2}{\xi+\frac{y_2}{h}}\right]d\xi,\nonumber
\end{eqnarray}
and
\begin{eqnarray}
  &&\cov(Z_{f_j},Z_{f_{\ell}})\nonumber\\
  &=&
    -\displaystyle{\lim_{r\downarrow 1}\frac{\kappa}{4\pi^2}\ointctrclockwise\limits_{|\xi_1|=1}
    \ointctrclockwise\limits_{|\xi_2|=1}
    \frac{f_j\left(\frac{1+h^2+2h\Re(\xi_1)}{(1-y_2)^2}\right)
      f_{\ell}\left(\frac{1+h^2+2h\Re(\xi_2)}{(1-y_2)^2}\right)}{(\xi_1-r\xi_2)^2}~d\xi_1d\xi_2}
  \label{cov1}\\
  &&  -\frac{(\beta_xy_1+\beta_yy_2)(1-y_2)^2}{4\pi^2
    h^2}\ointctrclockwise\limits_{|\xi_1|=1}
  \frac{f_j\left(\frac{1+h^2+2h\Re(\xi_1)}{(1-y_2)^2}\right)}
       {(\xi_1+\frac{y_2}{h})^2}d\xi_1
       \ointctrclockwise\limits_{|\xi_2|=1}\frac{f_{\ell}\left(\frac{1+h^2+2h\Re(\xi_2)}
         {(1-y_2)^2}\right)}{(\xi_2+\frac{y_2}{h})^2}d\xi_2.\nonumber
\end{eqnarray}
The proof of Theorem \ref{th:substitu2} is completed.

\subsection{Technical Lemmas}\label{sec:lemmas}

All the lemmas in this section assume that
the conditions of Theorem~\ref{th:substitu1} are satisfied.
For simplicity of proofs, we truncate the variables $X_{ij}$ as
$X_{ij}I_{(|X_{ij}|\leq n\eta_n)}$ where $\eta_n=o(1)$ because
truncation can't influence the proofs of theorems (see Page 183 of
Bai and Silverstein (2010)). For brevity of proofs, let $X_{ij}$
be the normalization of the truncated $X_{ij}I_{(|X_{ij}|\leq
n\eta_n)}$.

\begin{lem}\label{lem1} After truncation and normalization, for every $z\in{\mathbb C}^+=\{z: \Im(z)>0\}$, we have
$$
p(m_{n}^{(0)}-m^{(0)}_{N})\stackrel{a.s.}{\rightarrow}(1+z\underline{m}_{y})\cdot\frac{\underline{m}_{y}
+z\underline{m}'_{y_1}}{z\underline{m}_{y}}.
$$
\end{lem}

Proof. We have
\begin{equation}\label{eq3}
\underline{m}_{n}^{(0)}(z)=-\left(1-\frac{p}{n}\right)\cdot\frac{1}{z}+\frac{p}{n}m_{n}^0(z),\quad
\underline{m}_{N}^{(0)}(z)=-\left(1-\frac{p}{N}\right)\cdot\frac{1}{z}+\frac{p}{N}m_{N}^0(z)
\end{equation}
where $p/n\rightarrow {y_1}>0$. By (\ref{eq12}), we obtain
\begin{equation}\label{eq4}
\underline{m}'_{y_1}(z)=\frac{1}{\frac{1}{\underline{m}_{y}^2(z)}-{y_1}\int\frac{t^2}{(1+t\underline{m}_{y}(z))^2}dH(t)},\quad
{y_1}\int\frac{t}{1+t\underline{m}_{y}(z)}dH(t)=\frac{1+z\underline{m}_{y}(z)}{\underline{m}_{y}(z)}
\end{equation}
For brevity, $\underline{m}_{n}^{(0)}$,
$\underline{m}_{N}^{(0)}$, $\underline{m}_{y}(z)$ are
simplified as $\underline{m}_n^{(0)}$, $\underline{m}_{N}^{(0)}$
and $\underline{m}_{y}$. Using (\ref{eq1})-(\ref{eq2}), we
obtain
\begin{eqnarray*}
0&=&\frac{\underline{m}_{n}^{(0)}-\underline{m}_{N}^{(0)}}{\underline{m}_{n}^{(0)}\underline{m}_{N}^{(0)}}
-(\underline{m}_{n}^{(0)}-\underline{m}_{N}^{(0)})\frac{p}{n}\int\frac{t^2}{(1+t\underline{m}_{n}^{(0)})(
1+t\underline{m}_{N}^{(0)})}dH_p(t)\\
&&-\frac{p}{n(n-1)}\int\frac{t}{1+t\underline{m}_{N}^{(0)}}dH_p(t),
\end{eqnarray*}
that is,
\begin{equation}\label{eq5}
n(\underline{m}_{n}^{(0)}-\underline{m}_{N}^{(0)})=\frac{\frac{p}{N}\int\frac{t}
{1+t\underline{m}_{N}^{(0)}}dH_p(t)}
{\frac{1}{\underline{m}_{n}^{(0)}\underline{m}_{N}^{(0)}}-
\frac{p}{n}\int\frac{t^2}{(1+t\underline{m}_{n}^{(0)})(1+t\underline{m}_{N}^{(0)})}dH_p(t)}\rightarrow
\frac{{y_1}\int\frac{t}{1+t\underline{m}_{y}}dH(t)}
{\frac{1}{\underline{m}_{y}^2}-
{y_1}\int\frac{t^2}{(1+t\underline{m}_{y})^2}dH(t)}.
\end{equation}
By (\ref{eq3}), (\ref{eq4}) and (\ref{eq5}), we have \bqa
p(m_{n}^{(0)}-m^{(0)}_{N})&=&
n\underline{m}^{(0)}_{n}+\frac{n-p}{z}-\left((n-1)\underline{m}^{(0)}_{N}+\frac{n-1-p}z\right)\nonumber\\
&=&n(\underline{m}^{(0)}_{n}-\underline{m}^{(0)}_{N})+\underline m^{(0)}_{N}(z)+\frac{1}z\nonumber\\
&\rightarrow&
\frac{{y_1}\int\frac{t}{1+t\underline{m}_{y}}dH(t)}
{\frac{1}{\underline{m}_{y}^2}
-{y_1}\int\frac{t^2}{(1+t\underline{m}_{y})^2}dH(t)}+\frac{1+z\underline m_{y_1}(z)}{z}\nonumber\\
&=&\underline{m}_{y}^{'}\cdot\frac{1+z\underline{m}_{y}}{\underline{m}_{y}}+\frac{1+z\underline
m_{y_1}(z)}{z}
=(1+z\underline{m}_{y})\cdot\frac{\underline{m}_{y}+z\underline{m}_{y}'}{z\underline{m}_{y}}.\label{eq6}
\eqa Thus, Lemma \ref{lem1} is proved.   \eprf

In the sequel, we shall use Vatali lemma frequently. Let
$$
\bm{\gD}=\frac1n\sum_{j\ne k}\bgma_j\bgma_k^*.
$$
The normalization is by $1/n$ here instead of the
previously used $1/N$ but this difference does not affect the limits
calculated here.
We will derive the limit of
$\rtr(\bbA(z)-\bm{\gD})^{-1}-\rtr(\bbA^{-1}(z))$.
\begin{lem}\label{l1} After truncation and normalization, we have
$$\rE|\bgma_k^*\bbA^{-1}(z)\bgma_k-( 1+z\underline
m_{y_1}(z))|^2\le K n^{-1}$$ for every $z\in{\mathbb C}^+$ with a
constant $K$.
\end{lem}

\proof \ We have
$\bgma_k^*\bbA^{-1}(z)\bgma_k=\bgma_k^*\bbA^{-1}_k(z)\bgma_k\beta_k=1-\beta_k,$
where $\bbA_k(z)=\bbA(z)-\bgma_k\bgma_k^*$ and
$\beta_k=(1+\bgma_k^*\bbA_k^{-1}\bgma_k)^{-1}$. Therefore, By
(1.15) and (2.17) of Bai and Silverstein (2004), we have
$\rE|\bgma_k^*\bbA^{-1}(z)\bgma_k-(1+z\underline{m}_{y}(z))|^2=\rE|\beta_k+z\underline{m}_{y}(z)|^2\le
Kn^{-1}$.

\begin{cor}After truncation and normalization, we have
$$\rE\left|\bgma_k^*\bbA^{-2}(z)\bgma_k-\frac{d}{dz}(1+z\underline{m}_{y}(z))\right|^2\le
Kn^{-1}$$ for every $z\in{\mathbb C}^+$.
\end{cor}
\proof\ By Cauchy integral formula, we have
$$
\bgma_k^*\bbA^{-2}(z)\bgma_k=\frac1{2\pi
i}\oint_{|\zeta-z|=\Im(z)/2}\frac{\bgma_k^*\bbA^{-1}(\zeta)\bgma_k}{(\zeta-z)^2}d\zeta$$
and
$$
\frac{d}{dz}(1+z\underline{m}_{y}(z))=\frac1{2\pi
i}\oint_{|\zeta-z|=\Im(z)/2}\frac{1+\zeta\underline{m}_{y}(\zeta)}{(\zeta-z)^2}d\zeta.
$$

Then
$\rE\left|\bgma_k^*\bbA^{-2}\bgma_k-\frac{d}{dz}(1+z\underline{m}_{y}(z))\right|^2\le
Kn^{-1}$ follows from Lemma \ref{l1}.

\begin{lem}\label{l2} After truncation and normalization, we have
$\rE\left|\rtr{\bf A}^{-1}(z)\bm{\gD}\right|^2\le K n^{-1}$ for
every $z\in{\mathbb C}^+$. Especially for every $z\in{\mathbb
C}^+$,
$$
\rE|\rtr(\bbA^{-2}(z)\bm{\gD})|^2=\rE\left|\frac1n\sum_{j\ne
k\in\cU}\bgma_j^*\bbA^{-2}(z)\bgma_k\right|^2=O(n^{-1}).
$$
where $\cU=\{1,2,\cdots,n\}$.
\end{lem}

\proof \ We have
$\rtr\bbA^{-1}(z)\bm{\gD}=\frac1n\sum\limits_{j\ne
k\in\cU}\bgma_j^*\bbA^{-1}(z)\bgma_k=\frac1n\sum\limits_{j\ne
k\in\cU}\bgma_j^*\bbA_{jk}^{-1}(z)\bgma_k\beta_j\beta_{k(j)},$
where $\bbA_{jk}(z)=\bbA_k(z)-\bgma_j\bgma_j^*$ for $j\ne k$ and
$\beta_{k(j)}=(1+\bgma_k^*\bbA_{jk}^{-1}(z)\bgma_k)^{-1}$. We will
similarly define $\bbA_{ijk}(z)$ and $\beta_{k(ij)}$ for later
use. Then we obtain
$$
\begin{array}{lll}
\rE|\rtr(\bbA^{-1}(z)\bm{\gD})|^2&=&\rE\frac1n\sum\limits_{j_1\ne
k_1\in\cU}\bgma_{j_1}^*\bbA_{j_1k_1}^{-1}\bgma_{k_1}\beta_{j_1}\beta_{k_1(j_1)}
\frac1n\sum\limits_{j_2\ne k_2\in\cU}\overline{\bgma_{j_2}^*\bbA_{j_2k_2}^{-1}\bgma_{k_2}\beta_{j_2}\beta_{k_2(j_2)}}\\
&:=&\sum_{(2)}+\sum_{(3)}+\sum_{(4)},
\end{array}
$$
where the index $(\cdot)$ denotes the number of distinct integers
in the set $\{j_1,k_1,j_2,k_2\}$. By the facts that
$|\beta_{j}|\leq \frac{|z|}{\nu}$ and $\nu=\Im(z)$, we have
$$
\begin{array}{lll}
\sum_{(2)}&\le &\frac{2|z|^4}{n^2v^4}\sum\limits_{j\ne k\in\cU}\rE|\bgma_{j}^*\bbA_{jk}^{-1}\bgma_{k}|^2\\
&\leq&\frac{|z|^4}{\nu^4n^4}\sum\limits_{j\ne
k\in\cU}\rE\rtr(\bSi_x\bbA_{jk}^{-1}\bSi_x\overline{\bbA_{jk}^{-1}})\le
\frac{p}{n^2}\frac{|z|^4\|\bbT\|^2}{\nu^6}\le Kn^{-1},
\end{array}
$$
where $K$ is a constant. Moreover, we have
$$
\begin{array}{lll}
\sum_{(4)}&=&\frac{1}{n^2}\sum\limits_{j_1\ne k_1\ne j_2\ne
k_2\in\cU}
E\bgma_{j_1}^*\bbA^{-1}(z)\bgma_{k_1}\bgma_{j_2}^*\overline{\bbA^{-1}(z)}\bgma_{k_2}
\end{array}
$$
where
$$
\begin{array}{lll}
&&\bgma_{j_1}^*\bbA^{-1}(z)\bgma_{k_1}\\
&=&\beta_{j_1}\beta_{k_1(j_1)}\bgma_{j_1}^*\bbA_{j_1k_1}^{-1}\bgma_{k_1}\\
&=&\beta_{j_1}\beta_{k_1(j_1)}\Big[\bgma_{j_1}^*\bbA_{j_1k_1k_2}^{-1}\bgma_{k_1}-\beta_{k_2(j_1k_1)}\bgma_{j_1}^*\bbA_{j_1k_1k_2}^{-1}\bgma_{k_2}
\bgma_{k_2}^*\bbA_{j_1k_1k_2}^{-1}\bgma_{k_1}\Big]\\
&=&\beta_{j_1}\beta_{k_1(j_1)}\Big[\bgma_{j_1}^*\bbA_{j_1j_2k_1k_2}^{-1}\bgma_{k_1}-
\beta_{j_2(j_1k_1k_2)}\bgma_{j_1}^*\bbA_{j_1j_2k_1k_2}^{-1}\bgma_{j_2}
\bgma_{j_2}^*\bbA_{j_1j_2k_1k_2}^{-1}\bgma_{k_1}\\
&&-\beta_{k_2(j_1k_1)}\bgma_{j_1}^*\bbA_{j_1j_2k_1k_2}^{-1}\bgma_{k_2}
\bgma_{k_2}^*\bbA_{j_1j_2k_1k_2}^{-1}\bgma_{k_1}\\
&&+\beta_{k_2(j_1k_1)}\beta_{j_2(j_1k_1k_2)}
\bgma_{j_1}^*\bbA_{j_1j_2k_1k_2}^{-1}\bgma_{j_2}\bgma_{j_2}^*\bbA_{j_1j_2k_1k_2}^{-1}\bgma_{k_2}
\bgma_{k_2}^*\bbA_{j_1j_2k_1k_2}^{-1}\bgma_{k_1}\\
&&+\beta_{k_2(j_1k_1)}\beta_{j_2(j_1k_1k_2)}\bgma_{j_1}^*\bbA_{j_1j_2k_1k_2}^{-1}\bgma_{k_2}
\bgma_{k_2}^*\bbA_{j_1j_2k_1k_2}^{-1}\bgma_{j_2}\bgma_{j_2}^*\bbA_{j_1j_2k_1k_2}^{-1}\bgma_{k_1}\\
&&-\beta_{k_2(j_1k_1)}\beta^2_{j_2(j_1k_1k_2)}
\bgma_{j_1}^*\bbA_{j_1j_2k_1k_2}^{-1}\bgma_{j_2}\bgma_{j_2}^*\bbA_{j_1j_2k_1k_2}^{-1}\bgma_{k_2}
\bgma_{k_2}^*\bbA_{j_1j_2k_1k_2}^{-1}\bgma_{j_2}\bgma_{j_2}^*\bbA_{j_1j_2k_1k_2}^{-1}\bgma_{k_1}\Big]
\end{array}
$$

and \bqa
\beta_j&=&{b}_j-\beta_j{b}_j\epsilon_j={b}_j-{b}_j^2\epsilon_j+\beta_j{b}_j^2\epsilon^2_j\nonumber\\
\beta_{j(k)}&=&{b}_{j(k)}-\beta_{j(k)}{b}_{j(k)}\epsilon_{j(k)}={b}_{j(k)}-{b}_{j(k)}^2\epsilon_{j(k)}+\beta_{j(k)}{b}_{j(k)}^2\epsilon^2_{j(k)}
\label{eq45} \eqa with ${b}_j=\frac{1}{1+E\bgma_j^*{\bf
A}_j^{-1}(z)\bgma_j}$, $\epsilon_j=\bgma_j^*{\bf
A}_j^{-1}(z)\bgma_j-E\bgma_j^*{\bf A}_j^{-1}(z)\bgma_j$, and
$b_{j(k)}$ and $\epsilon_{j(k)}$ are similarly defined by
replacing $\bbA_j^{-1}(z)$ as $\bbA_{jk}^{-1}(z)$. By the same
manner, we can decompose $\bgma_{j_2}^*\bbA^{-1}(z)\bgma_{k_2}$
into similar 6 terms and then we will estimate the expectations of
the 36 products in the expansion of
$\bgma_{j_1}^*\bbA^{-1}(z)\bgma_{k_1}(\bgma_{j_2}^*\bbA^{-1}(z)\bgma_{k_2})^*$.

\noindent {\bf Case 1. Terms with at least six
$\bbA_{j_1j_2k_1,k_2}^{-1}(Z)$ in $\sum_{(4)}$.} We will prove
that these terms are bounded by $O(n^{-3})$. We shall use the fact
that all $\beta$-factors $\beta_j, \beta_{j(k)},
\beta_{k_2(j_1k_1)}, \beta_{j_2(j_1k_1k_2Z)}$ are bounded
$|z|/v\le K$. Let $\bbB=\bbA_{j_1j_2k_1,k_2}^{-1}(Z)$.  Say, for
the product of the two 6-th terms, its expectation is bounded by
\bqn
&&E\Big|(\bgma_{j_1}^*\bbB\bgma_{j_2}\bgma_{j_2}^*\bbB\bgma_{k_2}
\bgma_{k_2}^*\bbB\bgma_{j_2}\bgma_{j_2}^*\bbB\bgma_{k_1})(\bgma_{j_2}^*\bbB\bgma_{j_1}\bgma_{j_1}^*\bbB\bgma_{k_1}
\bgma_{k_1}^*\bbB\bgma_{j_1}\bgma_{j_1}^*\bbB\bgma_{k_2})^*\Big|\\
&\le&\left(\rE\Big|\bgma_{j_1}^*\bbB\bgma_{j_2}\bgma_{j_1}^*\bbB\bgma_{k_1}
\bgma_{k_2}^*\bbB\bgma_{j_2}\bgma_{j_2}^*\bbB\bgma_{k_1}\Big|^2
\rE\Big|\bgma_{j_2}^*\bbB\bgma_{j_1}\bgma_{j_2}^*\bbB\bgma_{k_2}
\bgma_{k_1}^*\bbB\bgma_{j_1}\bgma_{j_2}^*\bbB\bgma_{k_2}\Big|^2\right)^{1/2}.
\eqn  We have \bqn
&&\rE\Big|(\bgma_{j_1}^*\bbB\bgma_{j_2}\bgma_{j_1}^*\bbB\bgma_{k_1}
\bgma_{k_2}^*\bbB\bgma_{j_2}\bgma_{j_2}^*\bbB\bgma_{k_1})\Big|^2\\
&=&\frac1n\rE|(\bgma_{j_1}^*\bbB\bgma_{j_2}\bgma_{j_1}^*\bbB\bgma_{k_1}\bgma_{j_2}^*\bbB\bgma_{k_1})\Big|^2
\bgma_{j_2}^*\bbB^*\bSi_x\bbB\bgma_{j_2}\\
&\le&
\frac{K}{n^4v^2}\rE(\bgma_{j_2}^*\bbB^*\bSi_x\bbB\bgma_{j_2})^3
+\frac{K}{n^5}\sum_{i=1}^n\rE\Big|\bbe_i'\bm{\Gamma}^*\bbB\bgma_{j_2}\Big|^4\bgma_{j_2}^*\bbB^*\bSi_x\bbB\bgma_{j_2}\\
&\le&\frac{K}{n^4}\bigg[\|\bm{\Gamma}^*\bbB\bSi_x\bbB^*\bm{\Gamma}\|^3
+\frac1{n}\sum_{i=1}^n|\bbe_i'\bm{\Gamma}^*\bbB\bSi_x\bbB^*\bm{\Gamma}\bbe_i|^2
\cdot\|\bm{\Gamma}^*\bbB^*\bSi_x\bbB\bm{\Gamma}\|\bigg]
=O(n^{-4}), \eqn where $\bbe_i$ is the standard $i$-th unit
$p$-vector, i.e., its $i$-th entry is 1 and other $p-1$ entries 0.
In the last step of the above derivation, we have used facts that
$\rE|X_{ij_2}^6|\le \eta_n^2n\max \rE\rE|x_{ij}^4|=o(n)$ and
$\bbe_i'\bm{\Gamma}^*\bbB\bSi_x\bbB^*\bm{\Gamma}\bbe_i\le
\|\bSi_x\|^2/v^2$.

By similar approach, one can prove that the expectation of other
products with the number of $\bbB$ less than or equal to 6 are
bounded by $O(n^{-3})$.

\noindent {\bf Case 2. Terms with five
$\bbA_{j_1j_2k_1,k_2}^{-1}(Z)$ in $\sum_{(4)}$.} We shall use the
first expansion of $\beta_{j_1}$ and $\beta_{j_2}$ and then use
the bound
 bounded $|z|/v\le K$ for $\beta$'s. Then we can show that such terms are also bounded by $O(n^{-3})$.
 Say, for the product of the first term of $\bgma_{j_1}^*\bbA^{-1}(z)\bgma_{k_1}$ and the 6-th term
 of $\bgma_{j_2}^*\overline{\bbA^{-1}(z)}\bgma_{k_2}$, its expectation is bounded by
\bqn
&&\bigg|\rE(\beta_{j_1}\beta_{k_1(j_1)}\bgma_{j_1}^*\bbB\bgma_{k_1})(\beta_{j_2}\beta_{k_2(j_2)}\beta_{k_1(j_2k_2)}\beta_{j_1(j_2k_1k_2)}^2\bgma_{j_2}^*\bbB\bgma_{j_1}\bgma_{j_1}^*\bbB\bgma_{k_1}
\bgma_{k_1}^*\bbB\bgma_{j_1}\bgma_{j_1}^*\bbB\bgma_{k_2})^*\Big|\\
&=&\bigg|\rE\Big(\beta_{j_1}\beta_{k_1(j_1)}\beta_{j_2}\beta_{k_2(j_2)}\beta_{k_1(j_2k_2)}\beta_{j_1(j_2k_1k_2)}^2
-b_{j_1}b_{k_1(j_1)}b_{j_2}1_{k_2(j_2)}b_{k_1(j_2k_2)}b_{j_1(j_2k_1k_2)}^2\Big)\times\\
&&\ \
\bgma_{j_1}^*\bbB\bgma_{k_1}\bgma_{j_2}^*\bbB\bgma_{j_1}\bgma_{j_1}^*\bbB\bgma_{k_1}
\bgma_{k_1}^*\bbB\bgma_{j_1}\bgma_{j_1}^*\bbB\bgma_{k_2})^*\Big|\\
&\le&K\bigg(\rE\Big|\Big(\beta_{j_1}\beta_{k_1(j_1)}\beta_{j_2}\beta_{k_2(j_2)}\beta_{k_1(j_2k_2)}\beta_{j_1(j_2k_1k_2)}^2
-b_{j_1}b_{k_1(j_1)}b_{j_2}1_{k_2(j_2)}b_{k_1(j_2k_2)}b_{j_1(j_2k_1k_2)}^2\Big)\\
&&\ \ (\bgma_{j_1}^*\bbB\bgma_{k_1}\bgma_{j_2}^*\bbB\bgma_{j_1})
\Big|^2\rE\Big|(\bgma_{j_1}^*\bbB\bgma_{k_1}\Big|^4\Big|
\bgma_{j_1}^*\bbB\bgma_{k_2}\Big|^2\bigg)^{1/2}\le O(n^{-3}). \eqn
Here, we have used the fact that each term in the expansion of
$$\Big(\beta_{j_1}\beta_{k_1(j_1)}\beta_{j_2}\beta_{k_2(j_2)}\beta_{k_1(j_2k_2)}\beta_{j_1(j_2k_1k_2)}^2
-b_{j_1}b_{k_1(j_1)}b_{j_2}1_{k_2(j_2)}b_{k_1(j_2k_2)}b_{j_1(j_2k_1k_2)}^2\Big)$$
contains at leat one $\epsilon$ function. Then by the same
approach employed in Case 1, one can show that the bound is
$O(n^{-3})$.

\noindent {\bf Case 3. Terms with less than five
$\bbA_{j_1j_2k_1,k_2}^{-1}(Z)$ in $\sum_{(4)}$.} If there are four
$\bbA_{j_1j_2k_1,k_2}^{-1}(Z)$ in $\sum_{(4)}$, we need to further
expand the matrix $\bbA_{j_1}$ in $\epsilon_{j_1}$ as
$\bbA_{j_1j_2}^{-1}-\bbA_{j_1j_2}^{-1}\bgma_{j_2}\bgma_{j_2}^*\bbA_{j_1j_2}^{-1}\beta_{j_2(j_1)}$,
 expand $\bbA_{j_2}^{-1}=\bbA_{j_1j_2}^{-1}-\bbA_{j_1j_2}^{-1}\bgma_{j_1}\bgma_{j_1}^*\bbA_{j_1j_2}^{-1}\beta_{j_1(j_2)}$ in $\epsilon_{j_2}$,
 and then use the approach employed in Case 2 to obtain the desired bound.

 If the number is less than 4, we need to further expand the inverses of $\bbA$-matrices. The details are omitted. Finally, we obtain that
$$
\sum_{(4)}=O(\frac{1}{n}).
$$
Similarly, we have $$ \sum_{(3)}=O(\frac{1}{n}).
$$
Because
$\rtr(\bbA^{-2}\bm{\gD})=\frac{d}{dz}\rtr(\bbA^{-1}\bm{\gD})$,
then we have
$$
\rE|\rtr(\bbA^{-2}\bm{\gD})|^2=\rE\left|\frac1n\sum_{j\ne
k}\bgma_j^*\bbA^{-2}\bgma_k\right|^2=O(\frac{1}{n}).
$$
The lemma is proved.

\begin{lem}\label{l3}After truncation and normalization, we have
$\rtr(\bbA^{-2}\bm{\gD}\bbA^{-1}\bm{\gD})$ converges to
$(\underline{m}_{y}(z)+z\underline{m}_{y}'(z))(1+z\underline{m}_{y}(z))$
in $L_2$ for $z\in{\mathbb C}^+$.
\end{lem}

\proof \ Set
$\rtr\bbA^{-1}(z_1)\bm{\gD}\bbA^{-1}(z_2)\bm{\gD}=\frac1{n^2}\sum\limits_{i\ne
k, j\ne t}
\bgma_i^*\bbA^{-1}(z_1)\bgma_k\bgma_j^*\bbA^{-1}(z_2)\bgma_t
=Q_1+Q_2$ where
$$Q_1=\frac1{n^2}\sum_{j\not=k}^{n}\bgma_j^*\bbA^{-1}(z_1)\bgma_j\bgma_k^*\bbA^{-1}(z_2)\bgma_k
\quad\mbox{and}\quad Q_2=\frac1{n^2}\sum_{i\ne k, j\ne t\atop i\ne
j,{\rm or} k\ne
t}\bgma_i^*\bbA^{-1}(z_1)\bgma_k\bgma_j^*\bbA^{-1}(z_2)\bgma_t.
$$
By Lemma \ref{l1} and \ref{l2}, we obtain
$\rE|Q_1-(1+z\underline{m}_{y}(z_1))(1+z\underline{m}_{y}(z_2))|^2\leq
Kn^{-1}$ and $\rE|Q_2|^2=o(1).$ We thus have
$\rE|\rtr\bbA^{-1}(z_1)\bm{\gD}\bbA^{-1}(z_2)\bm{\gD}-(1+z\underline{m}_{y}(z_1))(1+z\underline{m}_{y}(z_2))|^2=o(1).$
Consequently, because
$\rtr\bbA^{-2}(z_1)\bm{\gD}\bbA^{-1}(z_2)\bm{\gD}=\frac{\partial
rtr\bbA^{-1}(z_1)\bm{\gD}\bbA^{-1}(z_2)\bm{\gD}}{\partial z_1}$,
then we have $
\rE|\rtr\bbA^{-2}(z_1)\bm{\gD}\bbA^{-1}(z_2)\bm{\gD}-\frac{\partial}{\partial
z_1}g(z_1)g(z_2)|^2=o(1). $ That is,
$\rtr\bbA^{-2}(z_1)\bm{\gD}\bbA^{-1}(z_2)\bm{\gD}$ converges to
$g(z_2)g'(z_1)$ in $L_2$ where $g(z)=1+z\underline{m}_{y}(z)$.
By setting $z_1=z_2=z$, we obtain
$\rtr(\bbA^{-2}\bm{\gD}\bbA^{-1}\bm{\gD})$ converges to
$g(z)g'(z)$ in $L_2$. Then Lemma \label{l3} is completed.
\begin{lem}\label{l4} After truncation and normalization, we have
$$\rtr(\bbA^{-1}\bm{\gD})^3(\bbA-\bm{\gD})^{-1}=g(z)\rtr((\bbA^{-1}\bm{\gD})^2(\bbA-\bm{\gD})^{-1})
+o_p(1)$$  for $z\in{\mathbb C}^+$.
\end{lem}
\proof\ We have \bqn
&&\rtr(\bbA^{-1}\bm{\gD})^3(\bbA-\bm{\gD})^{-1}
=E\frac1{n^3}\sum_{i\ne t, j\ne g\atop h\ne
s}\bgma_i^*\bbA^{-1}\bgma_j\bgma_g^*\bbA^{-1}\bgma_h\bgma_s^*
(\bbA-\bm{\gD})^{-1}\bbA^{-1}\bgma_t\nonumber\\
&=&\frac1{n^3}\sum_{i\ne t, j\ne g\atop i=j, h\ne
s}\bgma_i^*\bbA^{-1}\bgma_j\bgma_g^*\bbA^{-1}\bgma_h
\bgma_s^*(\bbA-\bm{\gD})^{-1}\bbA^{-1}\bgma_t\\
&& +\frac1{n^3}\sum_{i\ne t, j\ne g\atop i\ne j, h\ne
s}\bgma_i^*\bbA^{-1}\bgma_j\bgma_g^*\bbA^{-1}\bgma_h
\bgma_s^*(\bbA-\bm{\gD})^{-1}\bbA^{-1}\bgma_t\nonumber\\
&=&g(z)\frac1{n^2}\sum_{h\ne
s}\bgma_g^*\bbA^{-1}\bgma_h\bgma_s^*(\bbA-\bm{\gD})^{-1}\bbA^{-1}\bgma_t
+o_p(1)\\
&=&g(z)\rtr((\bbA^{-1}\bm{\gD})^2(\bbA-\bm{\gD})^{-1})+g(z)\frac1{n^2}\sum_{g=t\atop
h\ne s}\bgma_g^*\bbA^{-1}
\bgma_h\bgma_s^*(\bbA-\bm{\gD})^{-1}\bbA^{-1}\bgma_t
+o_p(1)\\
&=&g(z)\rtr((\bbA^{-1}\bm{\gD})^2(\bbA-\bm{\gD})^{-1}) +o_p(1).
\eqn Then by Lemma \ref{l3}, we have \bqn
\rtr(\bbA^{-1}\bm{\gD})^2(\bbA-\bm{\gD})^{-1}&=&\rtr(\bbA^{-1}\bm{\gD})^2(\bbA)^{-1}+\rtr(\bbA^{-1}\bm{\gD})^3(\bbA-\bm{\gD})^{-1}\\
&=&\rtr(\bbA^{-1}\bm{\gD})^2(\bbA)^{-1}+g(z)\rtr(\bbA^{-1}\bm{\gD})^2(\bbA-\bm{\gD})^{-1}+o_p(1)\\
&=&\frac{(1+z\underline{m}_{y}(z))(\underline{m}_{y}(z)+z\underline{m}_{y}'(z))}{1-g(z)}+o_p(1).
\eqn Hence, we obtain the following lemma.
\begin{lem}\label{thm2}After truncation and normalization, we have
\begin{eqnarray*}
&&tr{\bf A}^{-2}(z)\bm{\gD} +tr{\bf A}^{-1}(z)(\bm{\gD}{\bf
A}^{-1}(z))^2+tr\left({\bf A}(z)-\bm{\gD}\right)^{-1}(\bm{\gD}{\bf
A}^{-1}(z))^3\\
&=&\frac{(\underline{m}_{y}(z)+z\underline{m}_{y}'(z))(1+z\underline{m}_{y}(z))}{-z\underline{m}_{y}(z)}+o_p(1)
\end{eqnarray*}
 for $z\in{\mathbb C}^+$.
\end{lem}

\appendix

\section{Complements on the CLT Theorem~\protect\ref{th:newCLT}}
This appendix is intended to give more discussions on the CLT
Theorem~\ref{th:newCLT}.

\subsection{The special case where $\bSi\equiv I_p$}
In this special case, the  CLT for linear spectral statistics is
well-known since \citet{BS04} and the limiting mean and covariance
functions can be simplified significantly.  Here we report a
recent version proposed in \citet{WY13}. Then $H_p\equiv
\delta_{\{1\}}=H$ and the LSD $F^{y_1,H}$ becomes the standard
Mar\v{c}enko-Pastur distribution $F^{y_1}$ of index $y_1$.

\begin{prop} \label{thm201302}   
  Under the conditions of Theorem~\ref{th:newCLT} and
  assume moreover that $\bSi\equiv{\bf I}_{  p}$.
  Then the mean and
  covariance function of the Gaussian limit
  $(X_{f_1},\ldots,X_{f_k})$
  equal to
  \begin{eqnarray}
    \E[X_f]& =&   (\kappa-1)  I_1(f) + \beta_x  I_2(f) ~,
    \label{eq:E}\\
    \cov(X_{f},X_{g}) &=& \kappa J_1(f,g) + \beta_x
    J_2(f,g)~,
    \label{eq:cov}
  \end{eqnarray}
  where with $h_0=\sqrt y_1$,
  \begin{eqnarray}
    I_1(f) &=& \lim_{r\downarrow 1}\frac1{2\pi i}\oint_{|\xi|=1}f(|1+h_0\xi|^{2})\left[\frac{\xi}{\xi^{2}-r^{-2}}-\frac{1}{\xi}\right]d\xi    ~,\label{I1}\\
    I_2(f) &=&\frac{1}{2\pi i}\oint_{|\xi|=1}f(|1+h_0\xi|^{2})\frac{1}{\xi^{3}}d\xi~,\label{I2}\\
    J_1(f,g) &=&\lim_{r\downarrow 1}\frac{-1}{4\pi^{2}}\oint_{|\xi_{1}|=1}\oint_{|\xi_{2}|=1}
    \frac{f(|1+h_0\xi_{1}|^{2})g(|1+h_0\xi_{2}|^{2})} {(\xi_{1}-r\xi_{2})^{2}}d\xi_{1}d\xi_{2}   \label{J1}\\
    J_2(f,g)
   &=&-\frac{1}{4\pi^{2}}\oint_{|\xi_{1}|=1}\frac{f(|1+h_0\xi_{1}|^{2})}
   {\xi_{1}^{2}}d\xi_{1}\oint_{|\xi_{2}|=1}\frac{g(|1+h_0\xi_{2}|^{2})}{\xi_{2}^{2}}d\xi_{2}~.\label{J2}
  \end{eqnarray}
\end{prop}

\subsection{Comparison with the  CLT's in \cite{BS04} and
  \cite{PanZhou08}}

Compared to the  CLT in \citet{BS04}, Theorem
\ref{th:newCLT} removes
Gaussian-like 2nd-order and 4th-order moment conditions
and can then be applied to a broader range of
populations, e.g. non-Gaussian populations.
The new CLT relies on a new
condition that
{\em $\bm{\Gamma}^*\bm{\Gamma}$ is diagonal}.
This condition can hardly be relaxed as shown by the
following example.
\newcommand{\bbQ}{{\bf Q}}

The following example is for complex population. The example shows that although $\bm{\Sigma}_x=\diag[1,2,\cdots,1,2]$ is diagonal and Gaussian-like 4th moment condition exists,
there still is a counterexample that shows that the convergence of LSS of sample covariance matrices doesn't exist when $EX_{ij}^2\not=0$,
$\bm{\Gamma}^{*}\bm{\Gamma}$ is not diagonal and $\bm{\Gamma}$ is complex. The counterexample shows that
real $\bm{\Gamma}$ and diagonal $\bm{\Gamma}^{*}\bm{\Gamma}$ are unremoved for Theorem 2.1 when the Gaussian-like 2nd moment condition doesn't exist.
\begin{example}\label{exp1}
Let $p=2m$ and $\widetilde\bbT=\bm{\Gamma}^{*}\bm{\Gamma}=\bbU^{*}\bgL\bbU$ (i.e. $\bm{\Gamma}=\bgL^{1/2}\bbU$), where $\bm{\Sigma}_x=\bgL=\diag[1,2,\cdots,1,2]$ and
$$\bbU^{*}=\frac1{\sqrt2}\diag\left[\begin{pmatrix}1&e^{i\theta_{1m}}\cr e^{i\theta_{2m}}&-e^{i(\theta_{1m}+\theta_{2m})}\cr\end{pmatrix}
,\cdots,\begin{pmatrix}1&e^{i\theta_{1m}}\cr e^{i\theta_{2m}}&-e^{i(\theta_{1m}+\theta_{2m})}\cr\end{pmatrix}\right]$$
and $X_{ij}$s are i.i.d. and have a mixture distribution: with probability $\tau$ $X_{ij}\stackrel{D}=\frac{\sqrt{3}}{2}Y+\frac{i}2Z$ and probability $1-\tau$ $X_{ij}\stackrel{D}=\frac{\sqrt{3}}2W+\frac{i}2V$
where $Y, Z$ are i.i.d. standard normal and $W,V$ are i.i.d. and take values $\pm1$ with probability $\frac12$. Then, it is easy to verify that
$\rE X_{ij}=0$, $\rE |X_{ij}^2|=1$, $\rE X_{ij}^2=\frac12$ and $\rE|X_{ij}^4|=2\tau+\frac14$. Taking $\tau=\frac78$, we will have $\rE|X_{ij}^4|=2$.

Choose $f(x)=x$, then the random part of LSS is
$$
A_n(f)=\rtr{\bf S}_n^0-\rtr\widetilde\bbT=\frac1n\sjln (\bbx_j^*\bbx_j-\rtr\widetilde\bbT)
$$
where ${\bf X}_j=(x_{1j},\cdots,x_{pj})'$ and $\bbx_j=\bm{\Gamma}{\bf X}_j$. The variance of $A_n(f)$ is
$$
(E|X_{11}^4|-|EX_{11}^2|^2-2)\sum\limits_{j=1}^{p}\tilde{t}_{ii}+\rtr\widetilde\bbT^2+|\rE X_{11}^2|^2\rtr\widetilde\bbT\widetilde\bbT'
=\frac{6(\tau-1)m+5m}{n}+\frac{m(18+2\cos2\theta_{2m})}{16n}.
$$
Noting that $m/n\to y/2$, hence normalize LSS does not converge if we choose $\theta_{2m}$ such that $\cos^2\theta_{2m}$ does not have a limit.
\end{example}

The following example is for complex population. The example shows that although $\bm{\Sigma}_x=\diag[1,2,\cdots,1,2]$ is diagonal and Gaussian-like 2nd moment condition exists,
there still is a counterexample that shows that the convergence of LSS of sample covariance matrices doesn't exist when $E|X_{ij}^4|\not=2$
and $\bm{\Gamma}^{*}\bm{\Gamma}$ is not diagonal. The counterexample shows that
diagonal $\bm{\Gamma}^{*}\bm{\Gamma}$ is unremoved for Theorem 2.1 when the Gaussian-like 4th moment condition doesn't exist.
\begin{example}\label{exp2}
Let $p=2m$, $\bm{\Sigma}_x=\bm{\Gamma}\bm{\Gamma}^{*}$ with $\bm{\Gamma}=\bbU^{*}\bgL^{1/2}$ and $\widetilde\bbT=\bm{\Gamma}^{*}\bm{\Gamma}=\bbU^*\bgL\bbU$, where $\bgL=\diag[1,2,\cdots,1,2]$ and
$$\bbU^*=\diag\left[\begin{pmatrix}\cos\theta_m&\sin\theta_m\cr -\sin\theta_m&\cos\theta_{m}\cr\end{pmatrix}
,\cdots,\begin{pmatrix}\cos\theta_m&\sin\theta_m\cr -\sin\theta_m&\cos\theta_{m}\cr\end{pmatrix}\right]$$
and $X_{ij}$s are i.i.d. and have a mixture distribution: with probability $\tau$ $X_{ij}\stackrel{D}=\frac1{\sqrt{2}}(Y+iZ)$ and probability $1-\tau$ $X_{ij}\stackrel{D}=\frac1{\sqrt{2}}(W+iV)$
where $Y, Z,W$ and $V$ have the same distribution as given in Example \ref{exp1}. Then, it is easy to verify that
$\rE X_{ij}=0$, $\rE |X_{ij}^2|=1$, $\rE X_{ij}^2=0$ and $\rE|X_{ij}^4|=1+\tau$.

Choose $f(x)=x$, then the random part of LSS is
$$
A_n(f)=\rtr\bS_n^0-\rtr\widetilde\bbT=\frac1n\sjln (\bbx_j^*\bbx_j-\rtr\widetilde\bbT)
$$
where ${\bf X}_j=(X_{1j},\cdots,X_{pj})'$ and $\bbx_j=\bm{\Gamma}{\bf X}_j$. The variance of $A_n(f)$ is
$$
\frac1n\rtr\widetilde \bbT^2+\frac{\tau-1}{n}\sum_{i=1}^p \tilde{t}_{ii}^2=\frac{5m}n+\frac{m(\tau-1)(17-18\cos^2\theta_{m}\sin^2\theta_{m})}{n}.
$$
Again, the normalize LSS does not converge if we choose $\theta_{m}$ such that $\cos^2\theta_{m}$ does not have a limit.
\end{example}

The following example is for real population. The example shows that although $\bm{\Sigma}_x=\diag[1,2,\cdots,1,2]$ is diagonal and Gaussian-like 2nd moment condition exists,
there still is a counterexample that shows that the convergence of LSS of sample covariance matrices doesn't exist when real $X_{ij}$ satisfies $E|X_{ij}^4|\not=3$,
and $\bm{\Gamma}^{*}\bm{\Gamma}$ is not diagonal. The counterexample shows that
diagonal $\bm{\Gamma}^{*}\bm{\Gamma}$ is unremoved for Theorem 2.1 when the Gaussian-like 4th moment condition doesn't exist.
\begin{example}\label{exp3}
Choose $\bm{\Gamma}$ as sane as in Example \ref{exp2} and $X_{ij}$s are i.i.d. and their distribution is $\sqrt{3/5}$ times a $t$-distribution with degrees of freedom $5$. Then, it is easy to verify that
$\rE X_{ij}=0$, $\rE |X_{ij}^2|=1$, and $\rE|X_{ij}^4|=9$.

Again the variance of $A_n(f)$ is
$$
\frac2n\rtr\widetilde\bbT^2+\frac6n\sum_{i=1}^p\tilde{t}_{ii}^2=\frac{10m}n+\frac{6m(17-18\cos^2\theta_{m}\sin^2\theta_{m})}{n}.
$$
Hence, the normalize LSS does not converge if we choose $\theta_{m}$ such that $\cos^2\theta_{m}$ does not have a limit.
\end{example}


\end{document}